\documentclass[conference]{IEEEtran}
\usepackage{amsmath,amsfonts}
\usepackage{textcomp}
\usepackage{stfloats}
\usepackage{url}
\usepackage{verbatim}
\usepackage{graphicx}
\def\BibTeX{{\rm B\kern-.05em{\sc i\kern-.025em b}\kern-.08em
    T\kern-.1667em\lower.7ex\hbox{E}\kern-.125emX}}
\usepackage{balance}

\usepackage{amsmath}
\usepackage{amssymb}
\usepackage{booktabs}
\usepackage[american]{babel}
\usepackage{microtype}
\usepackage{xspace}
\usepackage{subcaption}
\usepackage[ruled,linesnumbered]{algorithm2e}
\usepackage{tcolorbox}
\usepackage{balance} 
\usepackage{enumitem} 
\usepackage{tabularx}
\usepackage{colortbl}
\usepackage{hyperref}
\usepackage{multicol}
\usepackage{multirow}
\usepackage{color}
\usepackage{pifont}
\usepackage{pdfpages}
\usepackage{tcolorbox}
\usepackage{mdframed}

\newcounter{mnote}
\newcommand{\specialcell}[2][c]{%
	\begin{tabular}[#1]{@{}c@{}}#2\end{tabular}}

\newcommand\approach{{DetectCodeGPT}\xspace}

\newtcolorbox{codebox}{
    colback=white, 
    colframe=black,
    boxrule=0.5pt, 
    boxsep=1pt,
    left=1pt,
    right=1pt,
    top=1pt,
    bottom=1pt,
}

\begin{document}


\title{Between Lines of Code: Unraveling the Distinct Patterns of Machine and Human Programmers}

\author{%
\IEEEauthorblockN{Yuling Shi}
\IEEEauthorblockA{
Shanghai Jiao Tong University\\
yuling.shi@sjtu.edu.cn}
\and
\IEEEauthorblockN{Hongyu Zhang}
\IEEEauthorblockA{
Chongqing University\\
hyzhang@cqu.edu.cn}
\and
\IEEEauthorblockN{Chengcheng Wan}
\IEEEauthorblockA{
East China Normal University\\
ccwan@sei.ecnu.edu.cn}
\and
\IEEEauthorblockN{Xiaodong Gu*}
\IEEEauthorblockA{
Shanghai Jiao Tong University\\
xiaodong.gu@sjtu.edu.cn}
\thanks{Xiaodong Gu is the corresponding author.}
}

\pagestyle{plain}

\maketitle

\begin{abstract}
  Large language models have catalyzed an unprecedented wave in code generation. While achieving significant advances, they blur the distinctions between machine- and human-authored source code, causing integrity and authenticity issues of software artifacts.
  Previous methods such as DetectGPT have proven effective in discerning machine-generated texts, but they do not identify and harness the unique patterns of machine-generated code. Thus, its applicability falters when applied to code. 
  In this paper, we carefully study the specific patterns that characterize machine- and human-authored code. Through a rigorous analysis of code attributes such as lexical diversity, conciseness, and naturalness, we expose unique patterns inherent to each source. 
  We particularly notice that the syntactic segmentation of code is a critical factor in identifying its provenance.
  Based on our findings, 
  we propose \approach, a novel method for detecting machine-generated code, which improves DetectGPT by capturing the distinct stylized patterns of code. Diverging from conventional techniques that depend on external LLMs for perturbations, \approach 
  perturbs the code corpus by strategically inserting spaces and newlines, ensuring both efficacy and efficiency. Experiment results show that our approach significantly outperforms state-of-the-art techniques in detecting machine-generated code.
  \footnote{Code available at \url{https://github.com/YerbaPage/DetectCodeGPT}}.
\end{abstract}

\begin{IEEEkeywords}
  machine-generated code detection, large language models, code generation, empirical analysis
\end{IEEEkeywords}

\section{Introduction}
$\let\thefootnote\relax\footnotetext{* Xiaodong Gu is the corresponding author}$

The advent of large language models (LLMs) such as Codex~\cite{chen2021evaluating} and ChatGPT~\cite{openai2022chatgpt} has revolutionized software engineering tasks such as code generation. 
Through extensive training on ultra-large code corpora~\cite{allal2023santacoder,nijkamp2023codegen2,zheng2023codegeex,roziere2023code}, LLMs acquire the ability to generate syntactically and functionally correct code, bringing a new era of efficiency and innovation in software creation, maintenance, and evolution.

While capable of generating human-like code, LLMs bring ambiguity of whether a software artifact is created by human or machine, causing integrity and authenticity issues in software development. This indistinction can lead to various challenges, such as misattribution of code ownership for bug triage and inflated assessments of developer workloads. Potential vulnerabilities in machine-generated code may go unnoticed due to the overreliance on its perceived robustness. 
The blending of human and machine efforts not only raises questions about the trustworthiness of the software but also threatens the integrity of development process, 
wherein the true authorship and the effort invested in creating software artifacts become obscured. Addressing these concerns is pivotal in maintaining a transparent and secure software development lifecycle.

Recently, there has been a growing research trend in detecting machine-generated texts~\cite{yang2023survey,zhao2023survey}. Perturbation-based methods like DetectGPT~\cite{mitchell2023detectgpt} have achieved state-of-the-art results in identifying machine-generated text. These methods employ likelihood score discrepancies between the original text and its various LLM-perturbed variants for detection, capturing the distinct patterns of machine-generated text that machines tend to prefer to a smaller set of phrases and expressions.
However, such detection methods tailored for natural language texts face challenges when applied to code, as code requires strict adherence to syntactic rules, while natural language can maintain coherence with more variation~\cite{lee2023who}. 
This situation highlights a significant gap in existing research: a lack of in-depth assessment of the intrinsic features of machine- and human-authored code, crucial for understanding the unique patterns of machine-generated code and devising effective detection methods.

In this paper, we conduct a comparative analysis of the distinct patterns between machine- and human-authored code from three aspects, including lexical diversity, conciseness, and naturalness. 
Through our analysis, we uncover that compared to human, machine tends to write more concise and natural code with a narrower spectrum of tokens and adhere to common programming paradigms, 
and the disparity is more pronounced in stylistic tokens such as the whitespace tokens.

Based on the findings, we propose a novel method called \approach for detecting machine-authored code. We extend the perturbation-based framework of DetectGPT by strategically inserting stylistic tokens, such as whitespace and newline characters, to capture the distinct patterns between machine- and human-authored code. This approach capitalizes on our observation that the disparity in coding styles is more pronounced in these stylistic tokens. By directly manipulating the code, \approach eliminates the need for an external pre-trained model, thereby enhancing both efficiency and effectiveness in the detection process.

To evaluate the effectiveness of \approach, we have conducted extensive experiments on two datasets across six code language models. The results demonstrate that \approach significantly outperforms 
the state-of-the-art methods by 7.6\% in terms of AUC.
Moreover, it proves to be a model-free and robust method against model discrepancies, making it viable for real-world applications with unknown or inaccessible source models.

Our contributions can be summarized as follows:
 \begin{itemize}[leftmargin=0.5cm]
  \item To our knowledge, we are the first to conduct a comprehensive and thorough analysis of the distinct patterns of LLM-generated code. Our study sheds light on essential insights that can further advance the utility of LLMs in programming.
     \item We propose a novel method for detecting machine-generated code by leveraging its distinct stylistic patterns.
     \item We extensively evaluate the \approach across a variety of settings and show the effectiveness of our approach.
 \end{itemize}

\section{Background}\label{sec_background}


\subsection{Large Language Models for Code}
Large language models~\cite{radford2018improving,radford2019language} based on Transformer~\cite{vaswani2017attention} decoder has achieved remarkable success in natural language processing tasks~\cite{ray2023chatgpt}. In the domain of code generation, Codex~\cite{chen2021evaluating} and AlphaCode~\cite{li2022competitionlevel} are pioneering works to train large language models on code. 
The training data often contain millions of code in different programming languages collected from open source repositories like GitHub~\cite{husain2020codesearchnet,kocetkov2022stack,gao2020pile}. 
Later advances to improve LLMs on code include designing new pretraining tasks like fill-in-the-middle~\cite{wang2021codet5,fried2022incoder,nijkamp2023codegen2,zheng2023codegeex} and also instruction fine-tuning~\cite{luo2023wizardcoder,roziere2023code}. Recent large language models pretrained on a mixture of programming and natural languages like ChatGPT~\cite{schulman2022chatgpt} and LLaMA~\cite{touvron2023llama} have also shown promising results on code generation tasks.


\subsection{Perturbation-Based Detection of Machine-Generated Text}\label{sec_detecting_machine_generated_text}
In the realm of machine-generated text detection, perturbation based method like DetectGPT~\cite{mitchell2023detectgpt} stands as the state-of-the-art technology~\cite{yang2023survey}.
In this section, we take DetectGPT as an example to illustrate the idea of perturbation-based detection methods. 
DetectGPT distinguishes between machine and human-generated text by analyzing the patterns in their probability scores~\cite{mitchell2023detectgpt}. 
The core idea is that when a text $x$ generated from a machine is subtly changed to $\tilde{x}$ through a perturbation process $q(\cdot|x)$ (e.g., MLM with T5~\cite{ahmad2021unified}), there is a sharper decline in its log probability scores $\log p_\theta(x)$ than that in human-generated text. This is because a machine-generated text is usually more predictable and tightly bound to the patterns it was trained on, leading to a distinct negative curvature in log probability when the text is perturbed.
By contrast, human-written texts are characterized by a rich diversity that reflects a blend of experiences and cognitive processes. As a result, it doesn't follow such predictable patterns, and its log probability scores $\log p_\theta(x)$ do not plummet as dramatically when similarly perturbed. 
Based on such discrepancy, we can define a \emph{likelihood discrepancy score} for each input code to measure the drop of log probability after perturbation.
\begin{equation}
  \mathbf{d}\left(x, p_\theta, q\right) \triangleq \log p_\theta(x)-\mathbb{E}_{\tilde{x} \sim q(\cdot \mid x)} \log p_\theta(\tilde{x})
\end{equation}
By inspecting these scores, we can detect the source of $x$. A significant drop indicates machine authorship and a smaller change suggests a human creator. This method effectively captures the more nuanced and variable nature of human-generated text compared to the more formulaic and patterned output of language models.

\section{Empirical Analysis}
\label{sec_empirical_analysis}

In this section, we conduct a comparative analysis of the distinct features of machine- and human-authored code.

\subsection{Study Design}\label{sec_study_design}

 To gain insights into distinctions between human and machine programmers, we consider three primary aspects that are relevant to coding styles, namely diversity, conciseness, and naturalness~\cite{zhang2008exploring,zhang2009discovering,hindle2016naturalness,albrecht1983software,rosenberg1997misconceptions}, which can be measured by specific metrics.

\subsubsection{Lexical Diversity}

Lexical diversity indicates the richness and variety of vocabulary present in a corpus. In the context of programming, this refers to the diversity in variable names, functions, classes, and reserved words. Analyzing lexical diversity offers a deeper understanding of the creativity, expressiveness, and potential complexity of code segments. There are four important empirical metrics in both natural and programming languages revealing the lexical diversity: token frequency, syntax element distribution, Zipf's law~\cite{zipf2016human} and Heaps' law~\cite{heaps1978information}.

\textbf{Token Frequency}
stands for the occurrence of distinct tokens in the code corpus. 
The attribute indicates the core vocabulary utilized by human and machine programmers, shedding light on their coding preferences and tendencies.



\begin{table}[t]
  \centering
  \caption{Studied categories of Python code tokens}
  \resizebox{\columnwidth}{!}{
  \begin{tabular}{|l|p{0.3\textwidth}|}
  \hline
  \textbf{Category} & \textbf{Tree-sitter Types} \\
  \hline
  keyword & def, return, else, if, for, while, $\ldots$ \\
  \hline
  identifier & identifier, type\_identifier \\
  \hline
  literal & string\_content, integer, true, false, $\ldots$ \\
  \hline
  operator & \textless, \textgreater, =, ==, +, $\ldots$ \\
  \hline
  syntactic symbol & \texttt{:}, \texttt{)}, \texttt{]}, \texttt{[}, \texttt{(}, \texttt{,}, \texttt{"}, \texttt{'}, \texttt{\{}, \texttt{\}}, \texttt{.} \\
  \hline
  comment & comment \\
  \hline
  whitespace & space, \textbackslash n \\
  \hline
  \end{tabular}}
  \label{tab:token_categories}
  \vspace{-0.5cm}
\end{table}

\textbf{Syntax Element Distribution}
refers to the proportion of syntax elements (e.g., keywords, identifiers) in the code corpus.
Understanding the distribution of syntax elements in code is akin to dissecting the anatomy of a language. It gives us a lens to view the nuances of coding style, the emphasis on structure, and the intricacies that distinguish human- and machine-authored code. 

To delve into the syntax element distribution, we analyze code with tree-sitter\footnote{\url{https://github.com/tree-sitter/tree-sitter}} and classify tokens into distinct categories, as detailed in Table \ref{tab:token_categories}. We then compute the proportion of each category in the code corpus.

\textbf{Zipf's and Heaps' Laws} were initially identified in natural languages~\cite{zipf2016human,heaps1978information}, and later verified in the scope of programming languages~\cite{zhang2008exploring,zhang2009discovering}.
Zipf's law states that the frequency value $f$ of a token is inversely proportional to its frequency rank $r$:
    $f \propto \frac{1}{r^\alpha}$,
where $\alpha$ is close to $1$~\cite{zipf2016human}. In programming languages, it states that a few variable names or functions are very commonly used across different scripts, while many others are rarely employed.
Heaps' Law characterizes the expansion of a vocabulary $V$ as a corpus $D$ increases in size: $V$ $\propto$ $D^{\beta}$, where $\beta$ $\in$ (0, 1) captures the rate of vocabulary growth relative to the size of the corpus. 

We investigate how closely machine-authored code aligns with Zipf's and Heaps' laws compared to human-authored code, which could reflect the models' ability to mimic human's lexical usage. 

\subsubsection{Conciseness}
Conciseness stands as a cornerstone attribute when characterizing code~\cite{barron1998minimum,albrecht1983software,rosenberg1997misconceptions}. The intricate balance of code conciseness directly influences readability, maintainability, and even computational efficiency.
We investigate two metrics that characterize code conciseness, namely, the number of tokens and the number of lines. 

\textbf{Number of tokens} gives us an indication of verbosity and complexity, showing the detailed composition of the code~\cite{barron1998minimum}. 

\textbf{Number of lines} helps us understand organizational choices, as spreading code across more lines can reflect a focus on readability and structure~\cite{rosenberg1997misconceptions}. 

\subsubsection{Naturalness}
The concept of code naturalness suggests that programming languages share a similar degree of regularity and predictability with natural languages~\cite{hindle2016naturalness}. This idea has been operationalized by employing language models to assess the probability of a specific token's occurrence within a given context. Under this framework, we inspect how ``natural'' machine-generated code is compared to human-written code.

\textbf{Token Likelihood and Rank} are two metrics that measure the naturalness of each token in the studied code corpus.
The token likelihood stands for the probability $p$ of a token $x$ under the model $p_\theta$, denoted as $p_\theta(x)$. The rank of a token $x$ is the position of $x$ in the sorted list of all tokens based on their likelihoods, denoted as $r_\theta(x)$. Both metrics evaluate how likely a token is preferred by the model~\cite{ippolito2020automatic,gehrmann2019gltr,solaiman2019release}. 
We calculate log scores on each token and then take the average to represent the whole code snippet as advised in~\cite{gehrmann2019gltr}.
To pinpoint the code elements that most significantly affect the score discrepancies, we also present the mean scores on different syntax element categories in Table~\ref{tab:token_categories} for comparison. 





\begin{table*}[ht]
  \centering
  \caption{Top 50 tokens from human- and machine-authored code from CodeLlama}
  \begin{tabular}{c|l|l}
    \toprule
    \textbf{Rank} & \textbf{Human-Authored Tokens} & \textbf{Machine-Authored Tokens} \\
    \midrule
    1--10   & \textbf{. ( ) = ' , : self "} [           & \textbf{. - , ( ) self : " ' if} \\
    11--20  & \textbf{] if return} in for not None 0 1 == & \textbf{= return} not [ def \textbf{raise} isinstance ] == path \\
    21--30  & else + is name \{ \} path data \textbf{raise} - & name 0 \textbf{\_\_class\_\_ \_\_name\_\_} None os \{ / \} \% \\
    31--40  & try * os len format get and True value isinstance & else \textbf{TypeError} str ` \_\_init\_\_ is $\textgreater$  // the \\
    41--50  & args key \% np i x kwargs except False or & in 1 ; value kwargs \#include + \_\_str\_\_ for \textbf{ValueError} \\
    \bottomrule
  \end{tabular}
  \label{tab:top_tokens}
  \vspace{-0.4cm}
\end{table*}

\subsection{Experimental Setup}\label{sec_experimental_setup_empirical_analysis}
We choose the state-of-the-art CodeLlama model~\cite{roziere2023code} to generate code. Limited by our computational resources, we use the version with 7B parameters. As for the decoding strategies, we adopt the top-$p$ sampling method~\cite{holtzman2019curious} with $p$=$0.95$ following~\cite{chen2021evaluating}.
The temperature $T$ is an important parameter controlling the diversity of the generated code~\cite{holtzman2019curious}. Since current LLMs on code are usually evaluated across different decoding temperatures~\cite{zheng2023codegeex,nijkamp2022codegen,chen2021evaluating,roziere2023code}, we generate code with $T=0.2$ and $T=1.0$ to capture the model's behavior under different settings. The maximum length of the generated code is set to 512 tokens based on the memory constraints and the length distribution of human-written code.
All experiments are conducted on 2 NVIDIA RTX 4090 GPUs with 24GB memory.

\subsection{Dataset Preparation}\label{sec_dataset_preparation}


To compare with human-authored code, we extract 10,000 Python functions randomly from the CodeSearchNet corpus~\cite{husain2020codesearchnet}, which is curated from a wide range of open-source GitHub projects.
We use the function signatures and their accompanying comments as prompts for the model as in~\cite{chen2021evaluating}. We also collect the corresponding bodies of these functions to represent human-written code.

While acknowledging that current models, including CodeLlama and even ChatGPT~\cite{liu2023your}, may not yet craft code of unparalleled quality for intricate tasks such as those in  CodeSearchNet~\cite{yetistiren2023evaluating,liu2023your}, the choice of this dataset is deliberate and insightful. 
Challenging the models against various real-world project code rather than simple programming problems, akin to those in the HumanEval~\cite{chen2021evaluating} or MBPP~\cite{austin2021program} dataset, offers a more representative assessment. It allows us to analyze the differences between human- and machine-authored code when faced with broader, practical applications. 

\subsection{Results and Analysis}\label{sec_results_and_analysis}
We present the results and analysis regarding each code attribute introduced in Section~\ref{sec_study_design}.

\subsubsection{Token Frequency}


Table~\ref{tab:top_tokens} lists the top 50 tokens from human- and machine-authored code when $T$=0.2. Due to space limit, we omit the results when $T$=1.0, which has a similar result.
From the results, we have several noteworthy observations:

\textit{Common Tokens}: Human- and machine-authored code shares a commonality in their usage of certain tokens, including punctuation marks such as ``.'', ``('', ``)'', and structural keywords such as ``if'', ``return'', and ``else''. This is because LLMs acquire foundational coding syntax after being trained on extensive human-written code corpora.

\textit{Error Handling}: Tokens associated with error handling like ``raise'' and ``TypeError'', are more prevalent in machine-authored code. This difference implies that machine programmers emphasize robustness and exception handling more explicitly.



\textit{Programming Paradigms}: Tokens indicating object-oriented programming like ``self'' and ``\_\_init\_\_'' are prominent in both human- and machine-authored code, which illustrates the model's training alignment with this paradigm. However, machine-authored code appears to favor more boilerplate tokens like ``\_\_class\_\_'' and also ``\_\_name\_\_'', which could stem from its training on diverse object-oriented codebase.


\begin{tcolorbox}[width=\linewidth, boxrule=0pt, sharp corners=all,
 left=2pt, right=2pt, top=2pt, bottom=2pt, colback=gray!20]
\textbf{Finding 1}: Machine-authored code pays more attention to exception handling and object-oriented principles than human, suggesting an emphasis on error prevention and adherence to common programming paradigms. 
\end{tcolorbox}

\subsubsection{Syntax Element Distribution}

\begin{figure}[t]
  \centering
  \includegraphics[width=0.4\textwidth, trim=0 20 0 0 clip]{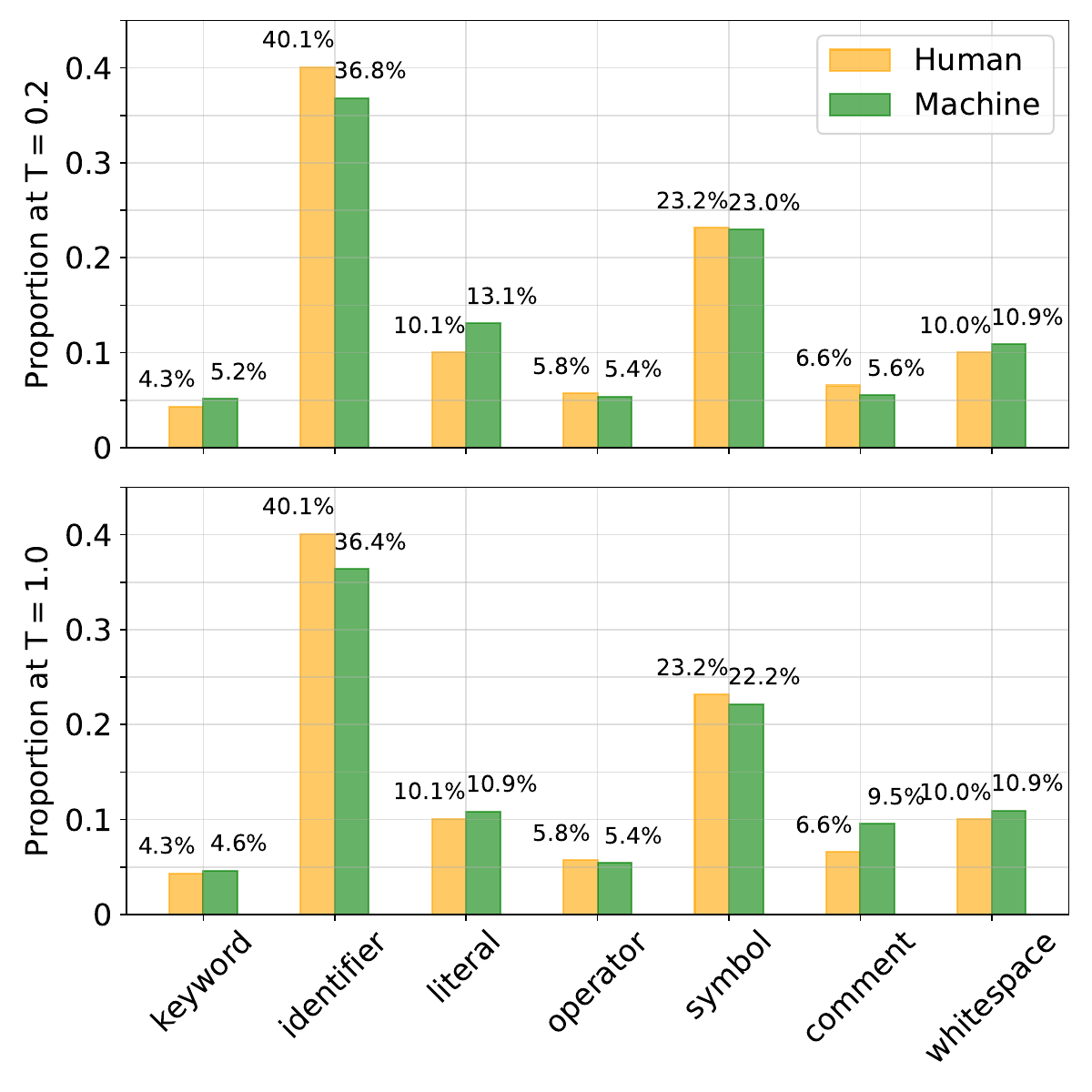}
  \caption{Syntax element distribution of the code corpus}
  \label{fig:token_category_distributions}
\end{figure}

The analysis of syntax element distributions, as visualized in Figure \ref{fig:token_category_distributions}, reveals intriguing insights into the coding conventions and stylistic nuances between human- and machine-authored code. The ``keyword'', ``operator'', ``syntactic symbols'', and ``whitespace'' proportions remain largely consistent between human and machine-authored code, across both temperatures. This consistency suggests that the foundational syntactical elements manifest similarly in both datasets. 
Delving deeper into the more nuanced discrepancies, a few categories emerge that underscore the differential preferences or tendencies of human and machine writers (statistical significant from Chi-square test with $p<0.01$):

\textit{Identifier}: The identifiers constitute a significantly lower proportion among machine-authored code across both temperatures, indicating that machine-authored code may have a more compact style with fewer identifiers.

\textit{Literal}: Machine-authored code consistently shows a slightly higher tendency towards using literals. Across both temperatures, the machine code exhibits an increase in the literal proportion compared to the human-written code. This suggests that machine-authored code may be more likely to process raw data directly. This could result from the machine's training on diverse data manipulation tasks.

\textit{Comment}: Machine-authored code has much more comments when $T$=1.0. This observation hints machine's increased emphasis on code documentation and explanation with higher temperatures, when it becomes less deterministic and more exploratory.

\begin{tcolorbox}[width=\linewidth, boxrule=0pt, sharp corners=all,
 left=2pt, right=2pt, top=2pt, bottom=2pt, colback=gray!20]
\textbf{Finding 2}: Machine-authored code tends to use fewer identifiers, more literals for direct data processing, and have more comments when the generation temperature grows.
\end{tcolorbox}

\subsubsection{Zipf's and Heaps' Laws}

\begin{figure*}[t]
  \begin{subfigure}{0.245\textwidth}
    \centering
    \includegraphics[width=\textwidth]{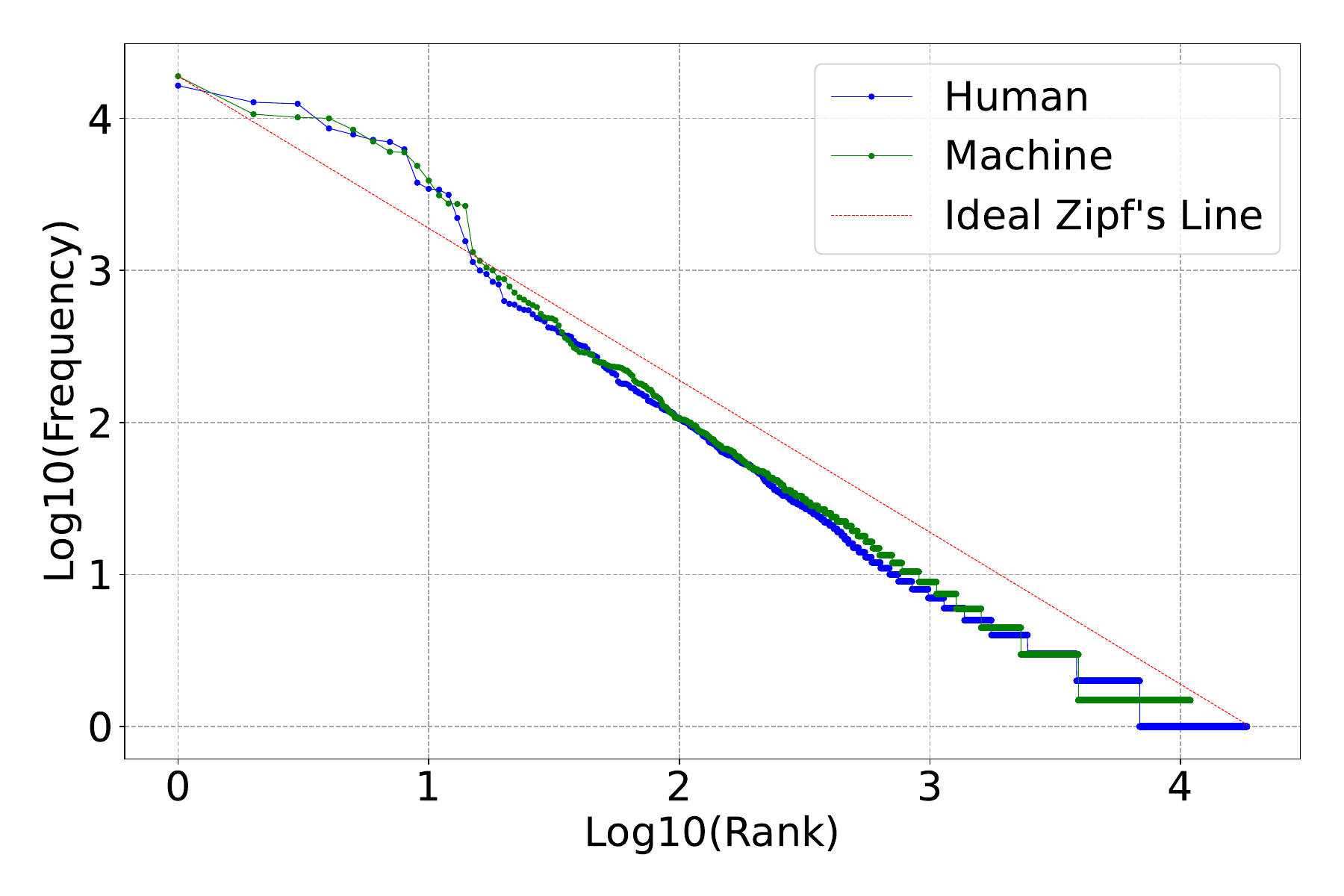}
    \caption{Zipf's Law ($T=0.2$)}
    \label{fig:combined_zipfs_tp0.2}
  \end{subfigure}
  \begin{subfigure}{0.245\textwidth}
    \centering
    \includegraphics[width=\textwidth]{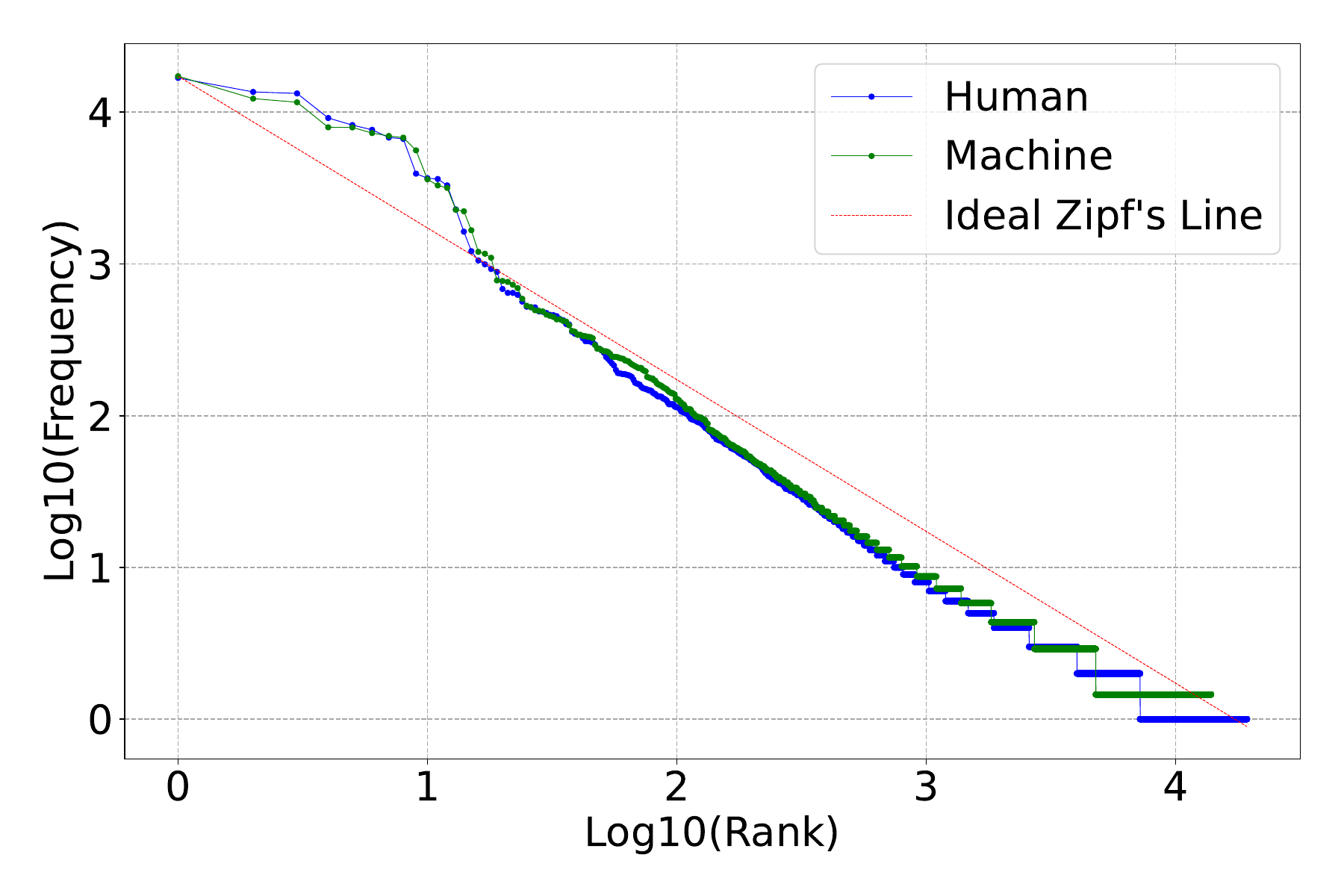}
    \caption{Zipf's Law ($T=1.0$)}
    \label{fig:combined_zipfs_tp1.0}
  \end{subfigure}
  \begin{subfigure}{0.245\textwidth}
    \centering
    \includegraphics[width=\textwidth]{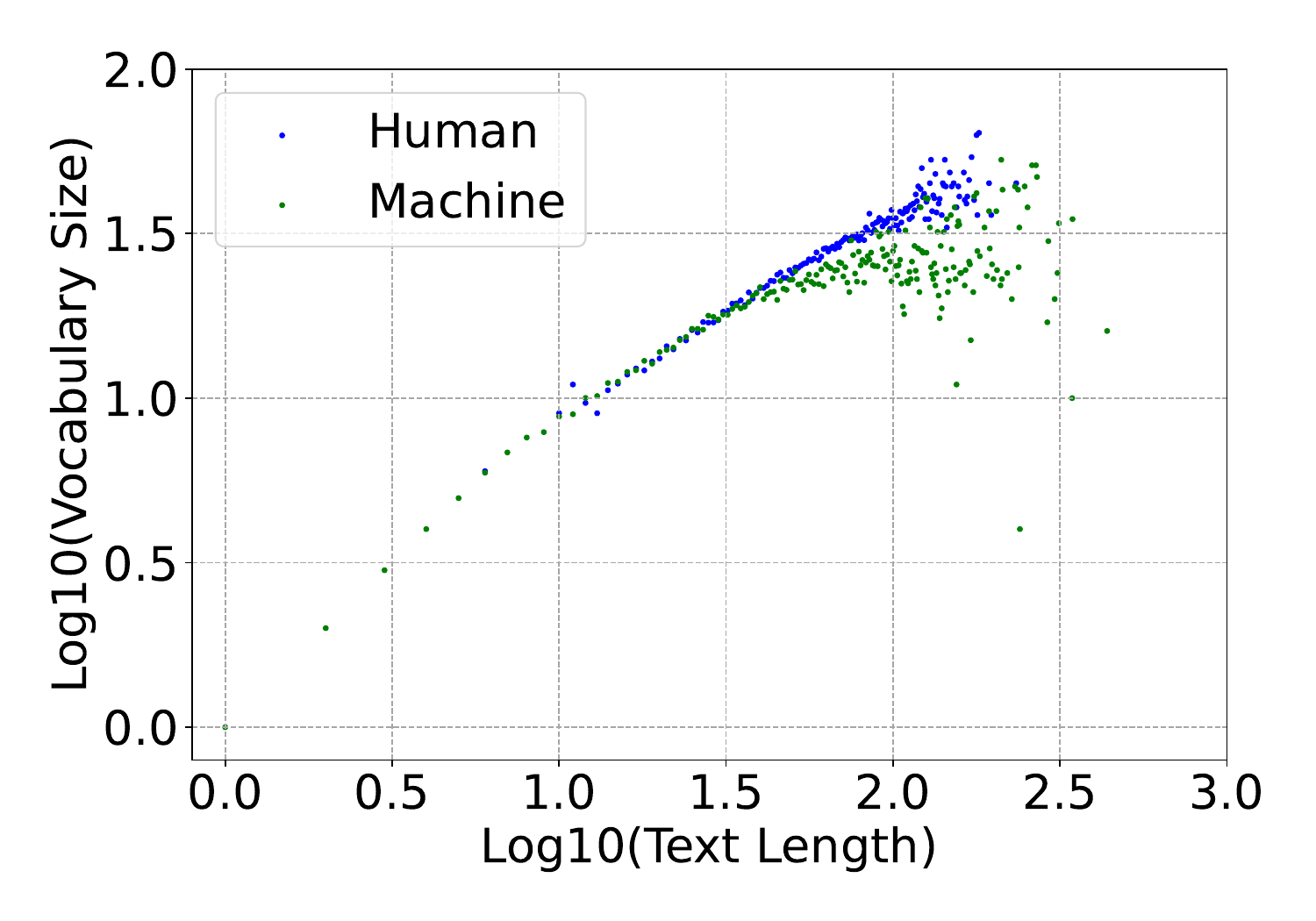}
    \caption{Heaps' Law ($T=0.2$)}
    \label{fig:combined_heaps_tp0.2}
  \end{subfigure}
  \begin{subfigure}{0.245\textwidth}
    \centering
    \includegraphics[width=\textwidth]{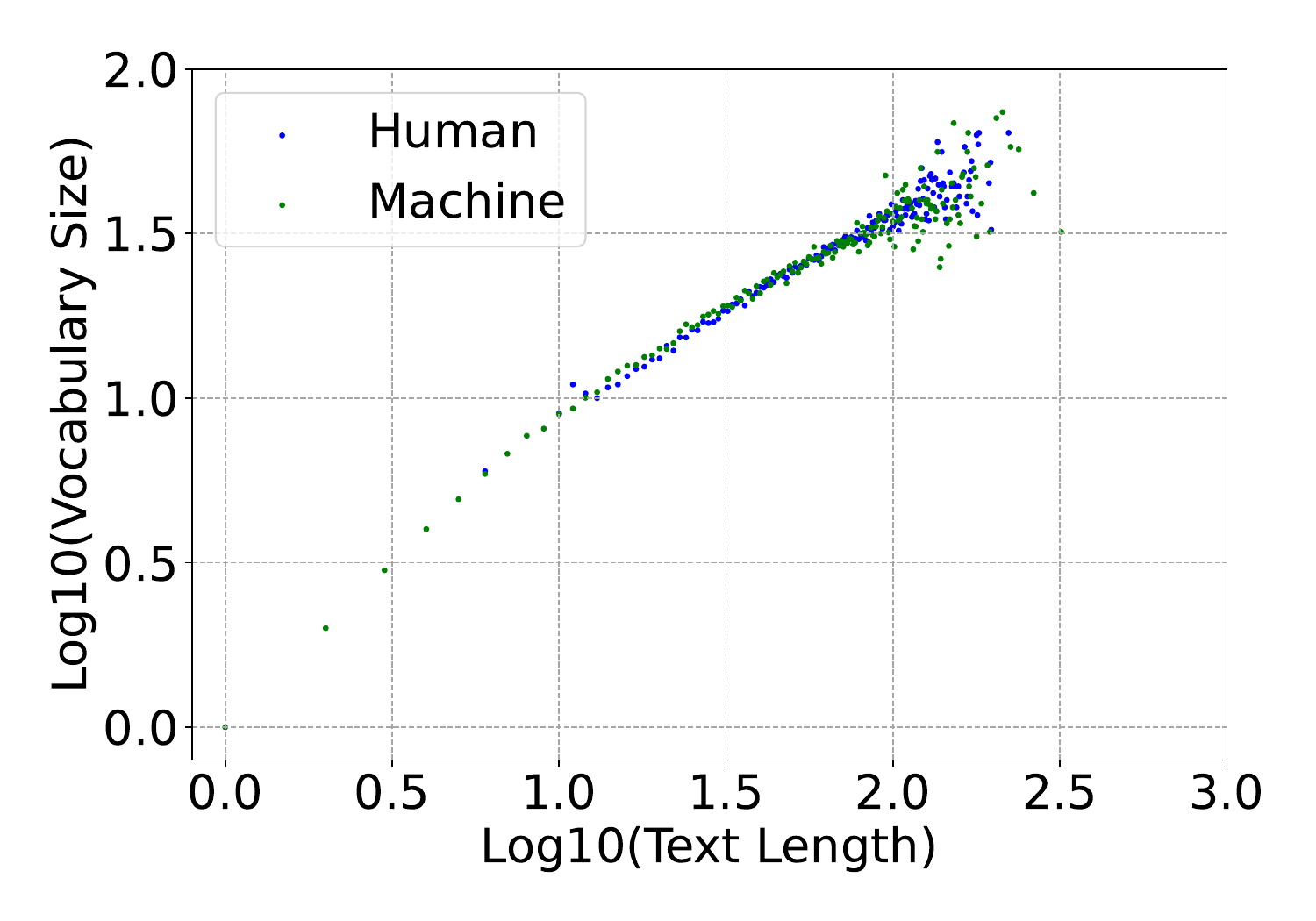}
    \caption{Heaps' Law ($T=1.0$)}
    \label{fig:combined_heaps_tp1.0}
  \end{subfigure}
  \caption{
    Comparison of Zipf's and Heaps' laws on machine- and human-authored code}
    \label{fig:combined_laws_comparison}
  \end{figure*}

Figure~\ref{fig:combined_laws_comparison} offers a comprehensive understanding of coding tendencies of both human and machine programmers. 
Starting with Zipf's law, Figure~\ref{fig:combined_zipfs_tp0.2} and \ref{fig:combined_zipfs_tp1.0} both delineate similar trends for human and machine programmers, corroborating the law's applicability. And we can observe machine's heightened proclivity towards tokens ranked between 10 and 100, especially at $T$=$0.2$.
Turning our attention to Heaps' law, the near-linear trends in Figure~\ref{fig:combined_heaps_tp0.2}-\ref{fig:combined_heaps_tp1.0} reaffirm the law's validity. 
Also, there's a noticeable shallowness in the slope for machine's code at $T$=$0.2$ revealing machine's decreased lexical diversity.


The obvious differences at $T$=$0.2$ can be ascribed to 
\textit{human creativity and variability}. The varied approaches and methodologies humans employ can lead to a diversified token usage within this range. Another plausible interpretation can be its \textit{risk aversion}. A machine, especially at lower temperatures, might be reverting to familiar patterns ensuring correctness in code generation. 
Additionally, certain patterns within the training data might have been overemphasized due to \textit{model overfitting}, leading the machine to a skewed preference.


\begin{tcolorbox}[width=\linewidth, boxrule=0pt, sharp corners=all,
  left=2pt, right=2pt, top=2pt, bottom=2pt, colback=gray!20]
 \textbf{Finding 3}: Machines demonstrate a preference for a limited spectrum of frequently-used tokens, whereas human code exhibits a richer diversity in token selection.
 \end{tcolorbox}

\subsubsection{Number of Tokens and Lines}

\begin{figure*}[t]
  \begin{subfigure}{0.24\textwidth}
      \centering
      \includegraphics[width=\textwidth]{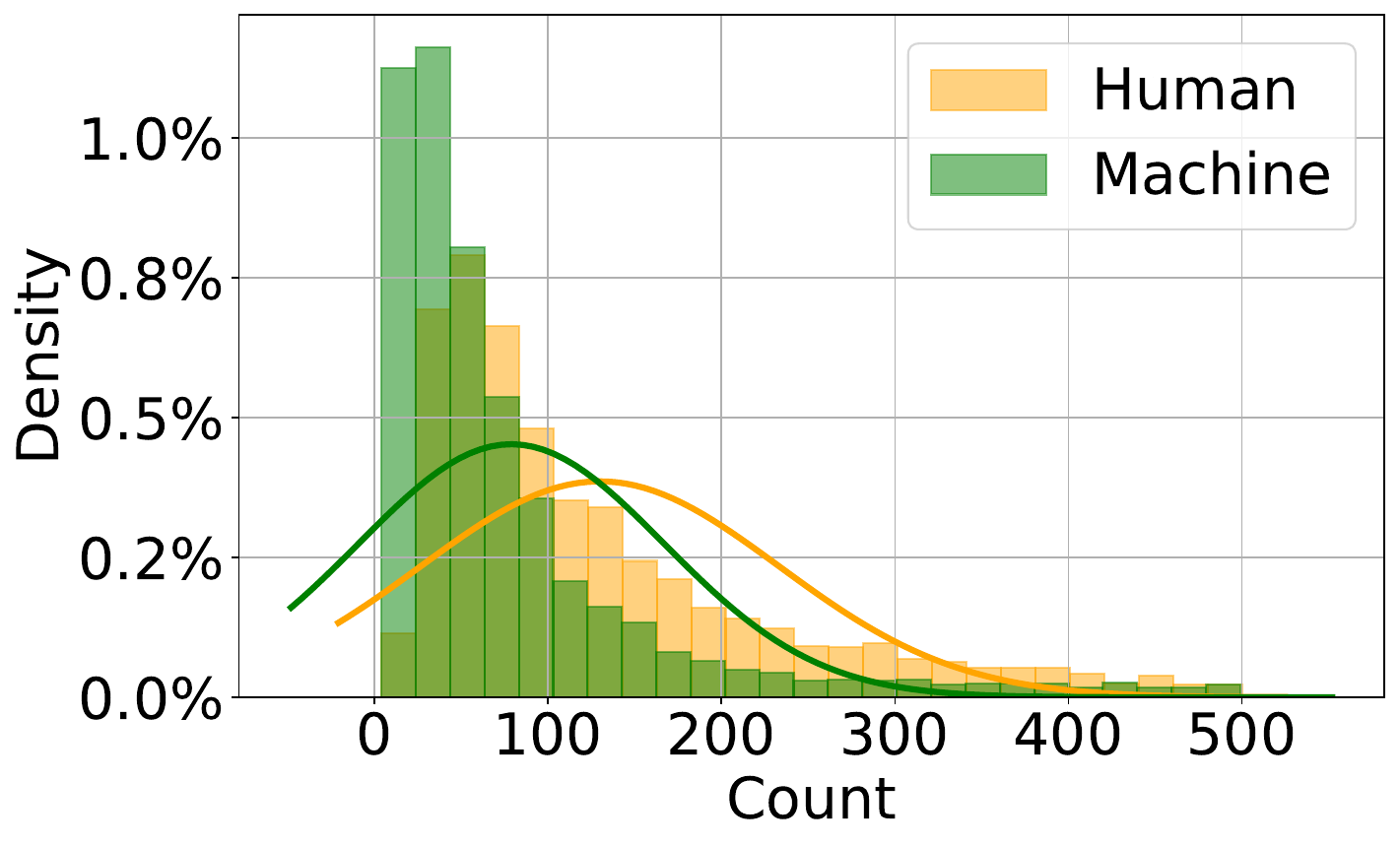}
      \caption{Number of tokens ($T=0.2$)}
      \label{fig:token_distribution_tp0.2}
  \end{subfigure}
  \begin{subfigure}{0.24\textwidth}
    \centering
    \includegraphics[width=\textwidth]{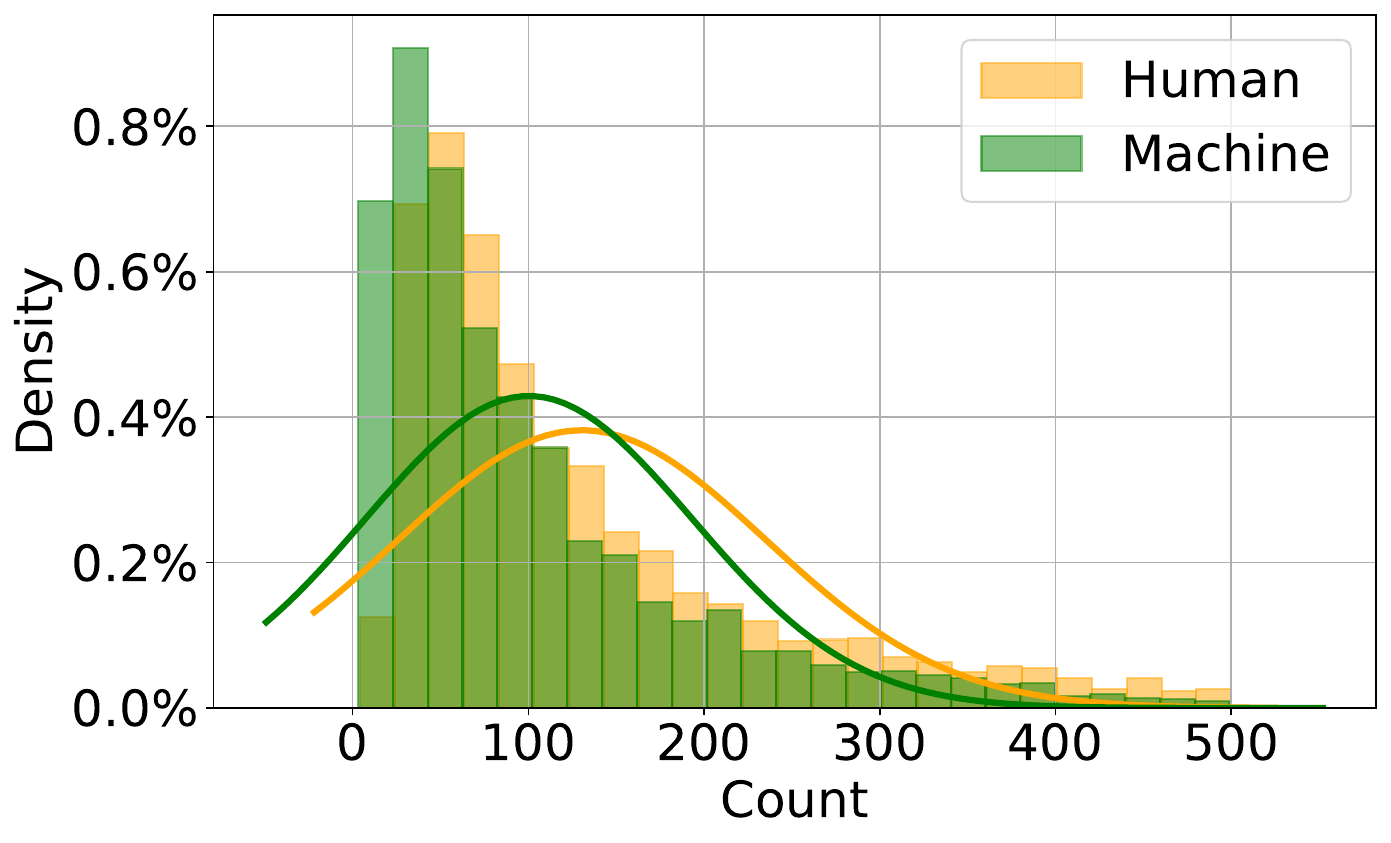}
    \caption{Number of tokens ($T=1.0$)}
    \label{fig:token_distribution_tp1.0}
  \end{subfigure}
  \begin{subfigure}{0.24\textwidth}
      \centering
      \includegraphics[width=\textwidth]{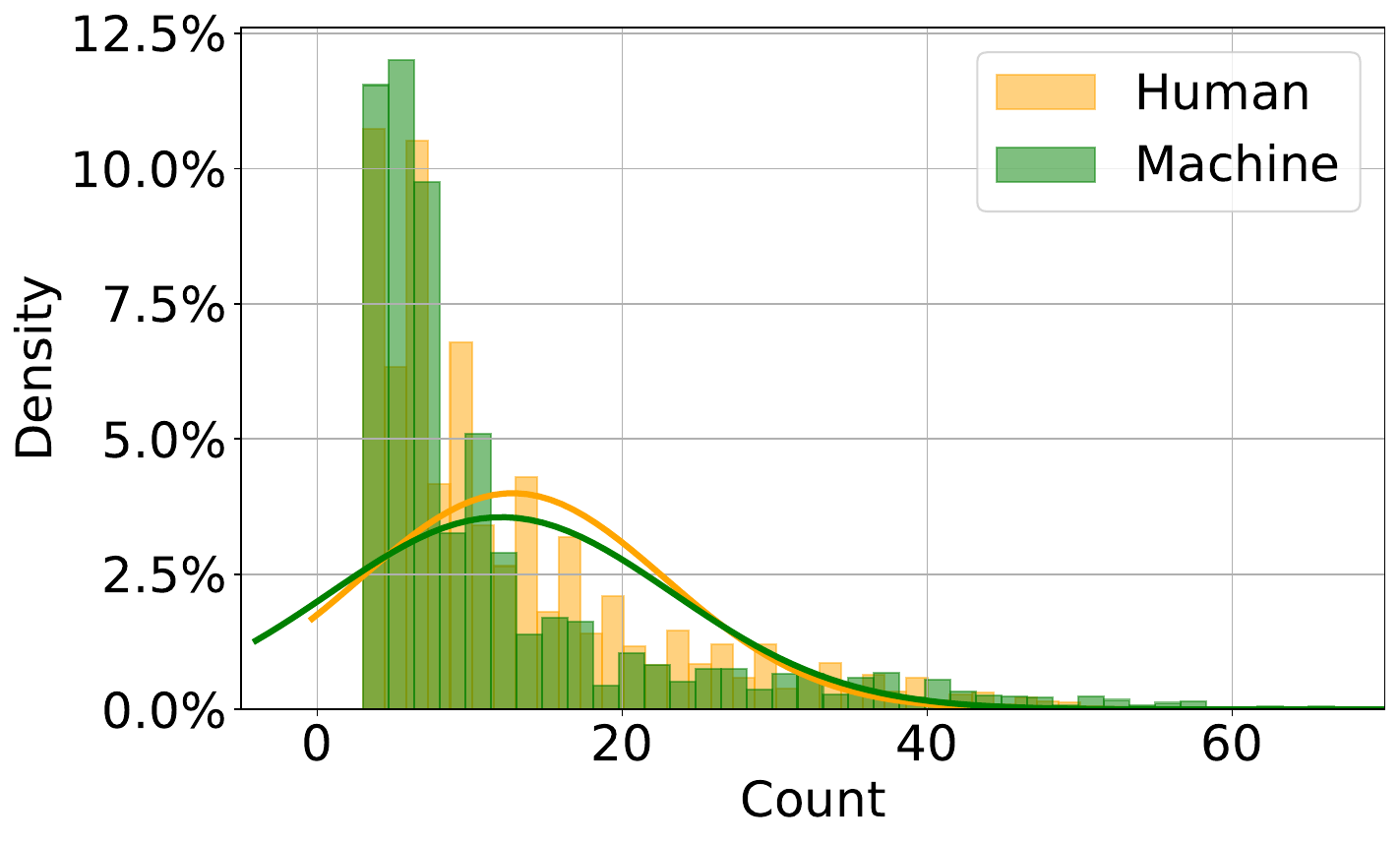}
      \caption{Number of lines ($T=0.2$)}
      \label{fig:line_distribution_tp0.2}
  \end{subfigure}
  \begin{subfigure}{0.24\textwidth}
      \centering
      \includegraphics[width=\textwidth]{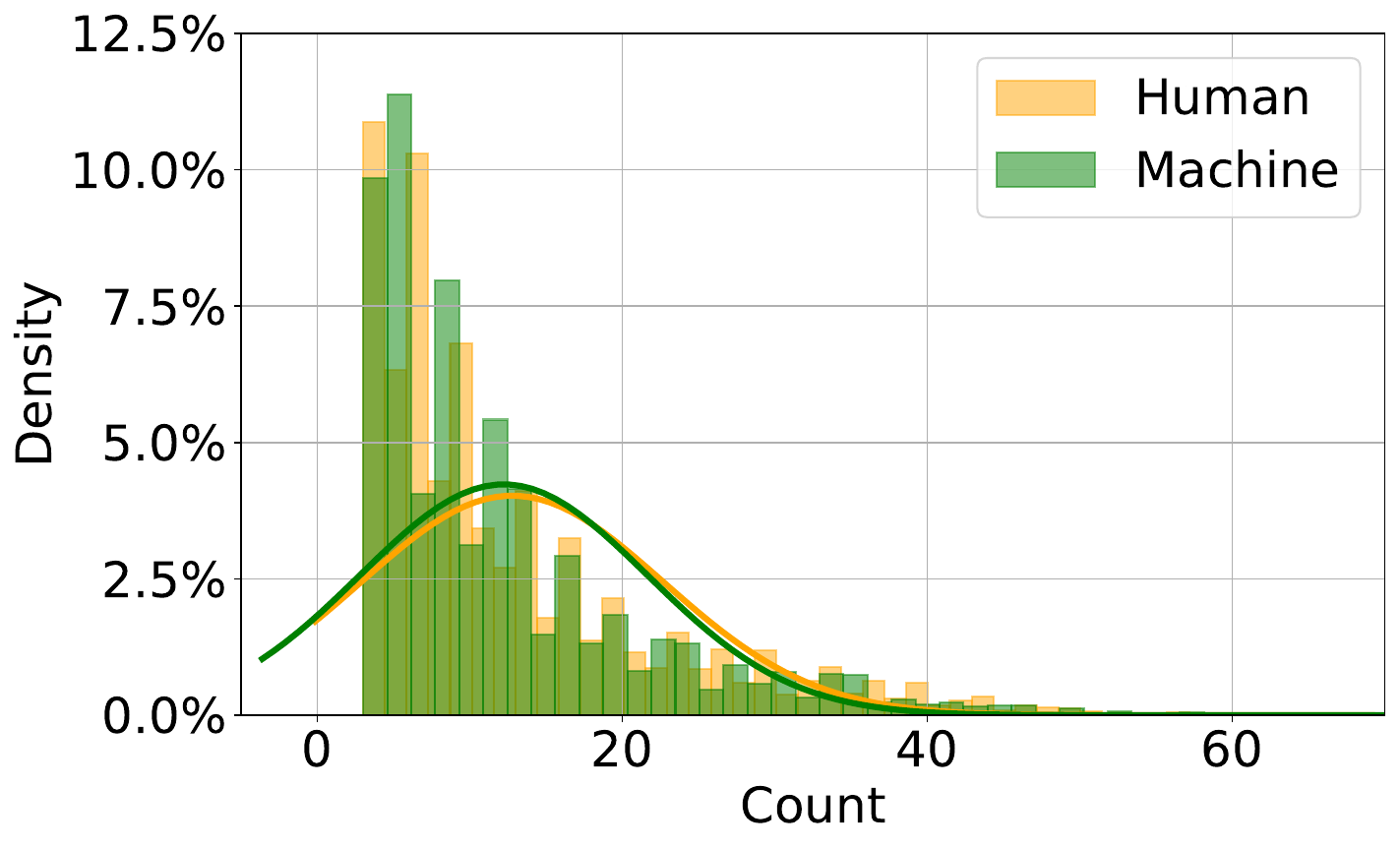}
      \caption{Number of lines ($T=1.0$)}
      \label{fig:line_distribution_tp1.0}
  \end{subfigure}
  \caption{
  Distribution of code length for machine- and human-authored code} 
  \label{fig:token_line_distribution}
\end{figure*}

Figure~\ref{fig:token_line_distribution} presents the distribution of code length under different settings. For the temperature setting of $0.2$, machine-authored code exhibits more conciseness, both in token and line numbers. 
As the temperature increases to $T=1.0$, we witness a convergence of distributions. The gap narrows, yet the machine's preference for relatively concise code persists. 
This reveals that higher temperatures induce more exploratory generative behavior in the model, leading to diverse coding styles. 

One could hypothesize several reasons for these observed patterns. The propensity for conciseness at lower temperatures may reflect LLM's training data, where probably concise solutions were more prevalent or deemed more ``correct''. On the flip side, human developers, often juggling multiple considerations like future code extensions, comments for peer developers, or even personal coding style, might craft lengthier solutions.
Furthermore, the narrowing of disparities at higher temperatures can be attributed to the model's increased willingness to explore varied coding styles. At higher temperatures, the LLM possibly mimics a broader spectrum of human coding patterns, capturing the essence of diverse coding habits and styles found in its training corpus.

\begin{tcolorbox}[width=\linewidth, boxrule=0pt, sharp corners=all,
 left=2pt, right=2pt, top=2pt, bottom=2pt, colback=gray!20]
\textbf{Finding 4}: Machines tend to write more concise code as instructed by their training objective, while human programmers tend to write longer code, reflective of their stylistic preferences. 

\end{tcolorbox}

\subsubsection{Token Likelihood and Rank}\label{sec_likelihood_rank}

\begin{figure}[t]
  \centering
  \includegraphics[width=0.9\linewidth, trim=0 20 0 0 clip]{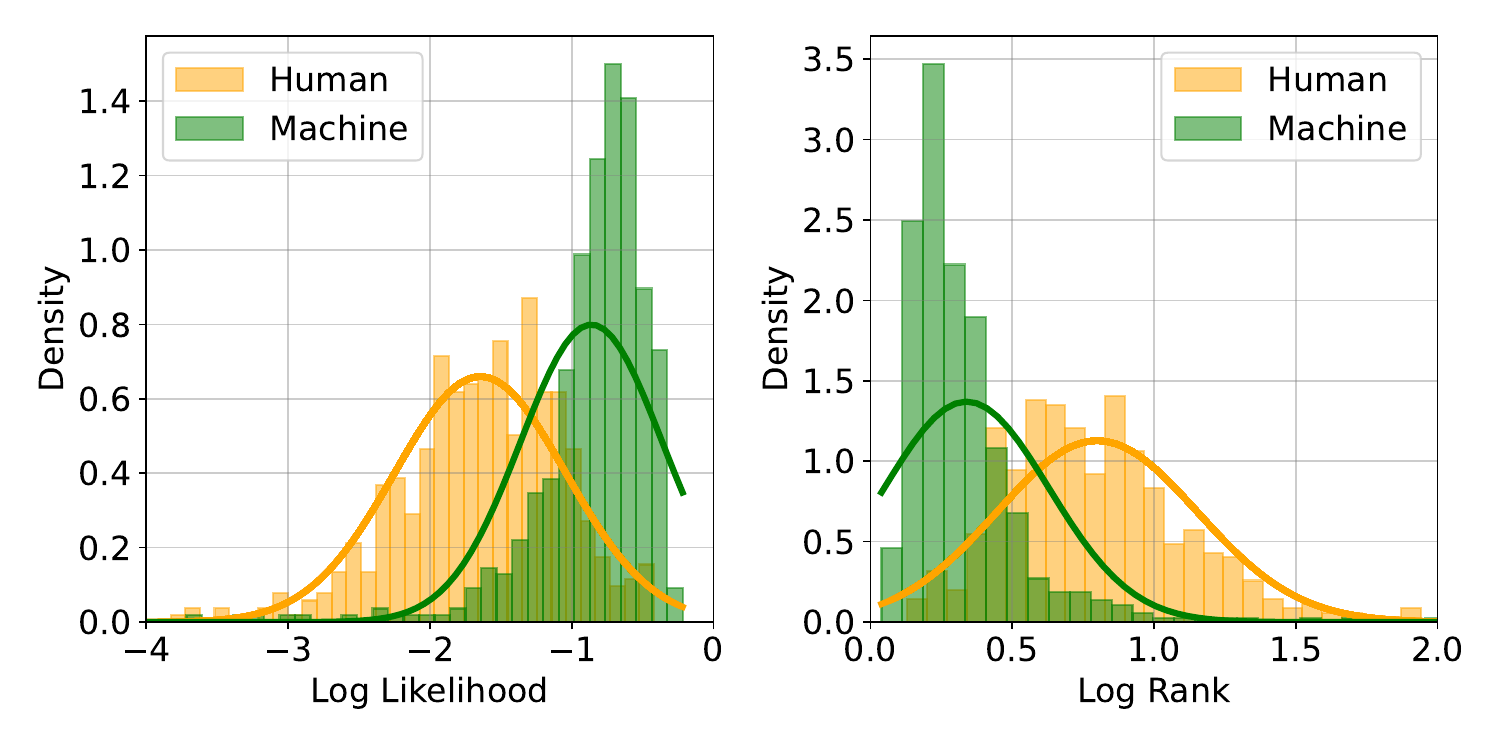}
  \caption{Distribution of naturalness scores}
  \label{fig:likelihood_rank_distribution}
  \vspace{-0.3cm}
\end{figure}

Figure~\ref{fig:likelihood_rank_distribution} shows that there is a great discrepancy of naturalness between machine- and human-authored code. Compared to human-authored code, the log likelihood scores of machine-authored code are mostly higher and the log rank scores are mostly lower, indicating that machine's code is more ``natural'' than human-written code. Such observation in source code is consistent with the findings in natural language~\cite{gehrmann2019gltr,ippolito2020automatic,mitchell2023detectgpt,solaiman2019release}. 

\begin{table}[t]
  \centering
  \caption{The naturalness of different categories of syntax elements. Statistical significance $p<0.001$.} 
  \resizebox{\linewidth}{!}{
  \begin{tabular}{l|ccc|ccc}
    \toprule
    \multirow{2}{*}{\bf Category} & \multicolumn{3}{c|}{\bf Log Likelihood} & \multicolumn{3}{c}{\bf Log Rank} \\ \cline{2-7}
     &  Machine  &  Human & $\Delta$ &  Machine  & Human & $\Delta$ \\
     \midrule 
     keyword     & -1.701 & -2.128 & 0.428  & 0.837  & 1.053 & 0.217 \\
     identifier  & -0.459 & -0.874 & 0.415  & 0.163  & 0.378 & 0.215 \\
     literal     & -0.506 & -1.364 & 0.858  & 0.152  & 0.630 & 0.479 \\
     operator    & -0.938 & -1.835 & 0.897  & 0.367  & 0.872 & 0.504 \\
     symbol      & -0.868 & -1.639 & 0.771  & 0.321  & 0.781 & 0.460 \\
     comment     & -1.503 & -3.028 & 1.525  & 0.608  & 1.610 & 1.002 \\
     whitespace  & -1.131 & -2.740 & \textbf{1.609}  & 0.429  & 1.441 & \textbf{1.012} \\
    \hline ALL   & -0.827 & -1.658 & 0.831  & 0.319 & 0.811 & 0.492 \\
    \bottomrule
  \end{tabular}
  }
  \label{tab:token_analysis}
  \vspace{-0.5cm}
\end{table}


Table~\ref{tab:token_analysis} summarises the comparison results in terms of each token category at $T=0.2$. 
An intriguing finding is that whitespace tokens stand out with the highest deviation of naturalness, surpassing even the combined use of all tokens. This highlights a distinctive aspect of coding styles: machines, trained on extensive datasets, typically generate code with regular, predictable whitespace patterns. Humans, however, influenced by individual styles and practices, exhibit a wider variety in their use of whitespaces. The distinct patterns in machine-generated whitespaces, therefore, point to an inherent variation in coding style between machine and human.

\begin{tcolorbox}[width=\linewidth, boxrule=0pt, sharp corners=all,
  left=2pt, right=2pt, top=2pt, bottom=2pt, colback=gray!20]
 \textbf{Finding 5}: Machine-authored code exhibits higher ``naturalness'' than human-authored code, and the disparity is more pronounced in tokens such as comments, and whitespaces, which are more reflective of individual coding styles.
\end{tcolorbox}




\section{Detecting Machine Generated Code}\label{sec_detection_method}
The empirical results suggest that machines tend to write more concise and natural code with a narrower spectrum of tokens, adhering to common programming paradigms, and the disparity is more pronounced in stylistic tokens such as whitespaces. 
This sparks a new idea for detecting machine-authored code: instead of perturbing arbitrary tokens, we focus on perturbing those stylistic tokens that best characterize the machine's preference. 
Based on this idea, we introduce \approach, a novel zero-shot method for detecting machine-authored code. 


\subsection{Problem Formulation}\label{sec_problem_definition}

We formulate the detection of machine-authored code as a classification task, which predict whether or not a given code snippet $x$ is produced by a source model $p_\theta$.
For this purpose, we transform $x$ to an equivalent form $\tilde{x}$ through a perturbation process $q(\cdot|x)$. We anticipate a sharper decline in its naturalness score if $x$ is written by an LLM. The key problems here are how to define the naturalness score and how to design the perturbation process. 
We introduce the naturalness score and the perturbation strategy $q(\cdot|x)$ in our approach in the following sections.

\subsection{Measuring Naturalness}
Previous methods usually use the log likelihood of tokens to measure the naturalness of machine-authored content~\cite{hindle2016naturalness,solaiman2019release}. However, the log rank of tokens shows better performance comparing the naturalness of machine- and human-authored text~\cite{mitchell2023detectgpt,gehrmann2019gltr}, because it offers a smoother and robust representation of token preference.

Unlike DetectGPT which directly calculates the log likelihood of tokens, we adopt the Normalized Perturbed Log Rank (NPR) score~\cite{su2023detectllm} to capture the naturalness. The NPR score is formally defined as:
\begin{equation}
\mathbf{NPR}\left(x, p_\theta, q\right) \triangleq \frac{\mathbb{E}_{\tilde{x} \sim q(\cdot \mid x)} \log r_\theta\left(\tilde{x}\right)}{\log r_\theta(x)},
\end{equation}
where $\log r_\theta(x)$ is the logarithm of the rank order of text $x$ sorted by likelihood under model $p_\theta$. In practice, $\mathbf{NPR}\left(x, p_\theta, q\right)$ has been demonstrated to be more accurate for differentiating text origins, outperforming the log likelihood discrepancy $\mathbf{d}\left(x, p_\theta, q\right)$~\cite{su2023detectllm}.

\subsection{Perturbation Strategy}


Our empirical study indicates that the whitespace tokens serve as an important indicator of machine's regularization and human's diversity, which points to an inherent variation in coding style.
Therefore, we propose an efficient and effective perturbation strategy with the following two types of perturbations below. Detailed explanations on the effectiveness of these perturbations are given in Section~\ref{sec_why_effective}.



\subsubsection{Space Insertion}
Let $ C $ represent the set of all possible locations to insert spaces in a code segment. We randomly select a subset $ C_s \subseteq C $ such that $ |C_s| = \alpha \times |C| $, where $\alpha \in [0,1]$ is a fraction representing the code locations. For each location $ c \in C_s $, we introduce a variable number of spaces, $ n_{\text{spaces}}(c) $, which is drawn from a Poisson distribution $ \mathcal{P}(\lambda_{\text{spaces}}) $. The Poisson distribution is chosen to simulate the randomness in human coding styles, similar to the random text infilling strategy in BART~\cite{lewis2020bart}. Mathematically, this can be represented as:
\begin{equation}
  n_{\text{spaces}}(c) \sim \mathcal{P}(\lambda_{\text{spaces}}).
\end{equation}

\subsubsection{Newline Insertion}
We split the code into lines and obtain a set $ L $ of lines. A subset $ L_n \subseteq L $ is then chosen randomly, where $ |L_n| = \beta \times |L| $, with $ \beta \in [0,1]$ denoting the proportion of the line locations. For each line $ l \in L_n $, we introduce a variable number of newlines, $ n_{\text{newlines}}(l) $, also sampled from a Poisson distribution $ \mathcal{P}(\lambda_{\text{newlines}}) $:

\begin{equation}
    n_{\text{newlines}}(l) \sim \mathcal{P}(\lambda_{\text{newlines}}).
\end{equation}

\begin{algorithm}[ht]
  \caption{\approach: Machine-Generated Code Detection with Stylized Code Perturbation}
  \label{alg_detection}
  \small
  \KwData{code $x$, source model $\mathcal{M}$, number of perturbations $k$, decision threshold $\epsilon$, parameters $\alpha$, $\beta$, $\lambda_{\text{spaces}}$, and $\lambda_{\text{newlines}}$}
  
  \For{$i \leftarrow 1$ \KwTo $k$}{
      \tcp{Random decision for type of perturbation}
      $ p \sim \mathcal{U}(0, 1) $\;
      \eIf{$p \leq 0.5$}{
          \tcp{Spaces Insertion}
          Let $ C $ represent all possible locations to insert spaces in $x$\;
          Select $ C_s \subseteq C $ such that $ |C_s| = \alpha \times |C| $\;
          \For{each location $ c \in C_s $}{
              $ n_{\text{spaces}}(c) \sim \mathcal{P}(\lambda_{\text{spaces}}) $\;
              Insert $ n_{\text{spaces}}(c) $ spaces at location $ c $ in $x$\;
          }
      }{
          \tcp{Newlines Insertion}
          Split the perturbed code $x$ into a set $ L $ of lines\;
          Select $ L_n \subseteq L $ such that $ |L_n| = \beta \times |L| $\;
          \For{each line $ l \in L_n $}{
              $ n_{\text{newlines}}(l) \sim \mathcal{P}(\lambda_{\text{newlines}}) $\;
              Insert $ n_{\text{newlines}}(l) $ newlines after line $ l $ in $x$\;
          }
      }
      Store the perturbed code as $\tilde x_i$\;
  }
  
  Estimate NPR: $\text{NPR}_x \gets \text{NPR}\left(x, p_\theta, q\right) - \frac{1}{k}\sum_i\text{NPR}\left(\tilde x_i, p_\theta, q\right)$\;
  
  \eIf{$\text{NPR}_x > \epsilon$}{
      \Return \texttt{true} \tcp*[r]{Probably machine-authored}
  }{
      \Return \texttt{false} \tcp*[r]{Probably human-authored}
  }
  
\end{algorithm}

We randomly choose one type of perturbation to the code snippet $x$ to generate a set of perturbed samples $\tilde x_i$ for $i \in [1, k]$, where $k$ is the number of perturbations. Through this step, we instill randomness at a granular stylistic level, thereby amplifying the perturbation's efficacy. Our perturbation strategy introduces several distinct advantages over the conventional methods~\cite{mitchell2023detectgpt,su2023detectllm} using MLM to perturb the code, which will be discussed in Section~\ref{sec_strength}.

Algorithm \ref{alg_detection} summarizes the entire workflow of \approach. Our algorithm harnesses stylized code perturbation to differentiate between human- and machine-authored code. At the core of our approach is the strategic insertion of spaces (Lines 4-9) and newlines (Lines 11-16) in code, a process that simulates the inherent randomness in human coding styles. The algorithm operates by generating perturbed versions of the code and then evaluating their NPR scores (Lines 20-21) with respect to the source model $\mathcal{M}$. 

The threshold parameter $\epsilon$ in Line 22, pivotal for making the detection decision, offers flexibility in catering to different application scenarios. By adjusting $\epsilon$, users can balance between false positives and false negatives, tailoring the detection sensitivity according to the specific needs of the deployment context.


\section{Evaluation}\label{sec_evaluation}

We conduct experiments to evaluate the effectiveness of \approach, aiming to answer the following research questions.

\begin{itemize}[leftmargin=*]
    \item \textbf{RQ1:} How effectively does our method distinguish between machine-generated and human-written code?
    \item \textbf{RQ2:} To what extent do individual components influence the overall performance of our method?
    \item \textbf{RQ3:} What is the impact of varying the number of perturbations on the detection performance?
    \item \textbf{RQ4:} How effective is our method in cross-model code detection?
\end{itemize}

\subsection{Datasets}\label{sec_dataset_and_metrics}
We carefully use a different split of the \textit{CodeSearchNet} dataset~\cite{husain2020codesearchnet} from the one used in the empirical study for evaluation. We select Python code from \textit{The Stack}~\cite{kocetkov2022stack} as another evaluation dataset. Similar to \textit{CodeSearchNet}, \textit{The Stack} provides code from a variety of open-source projects representative of real-world scenarios. We use a parsed and filtered version~\cite{manh2023vault} of this dataset and also concatenate the function definitions with their corresponding comments as prompts as in~\cite{chen2021evaluating}. For each combination of dataset and model, we sample 500 human and machine code pairs for evaluation. The maximum length of code is trimmed to 128 tokens.



\subsection{Studied Models}
We investigate machine-generated code by a diverse array of advanced LLMs, including Incoder~\cite{fried2022incoder}, Phi-1~\cite{gunasekar2023textbooks}, StarCoder~\cite{li2023starcoder}, WizardCoder~\cite{luo2023wizardcoder}, CodeGen2~\cite{nijkamp2023codegen2} and CodeLlama~\cite{roziere2023code}.
We obtain their checkpoints from Huggingface~\footnote{\url{https://huggingface.co/models}} with different parameter sizes (1B-7B).

\subsection{Evaluation Metric}

Following prior works~\cite{mitchell2023detectgpt,su2023detectllm}, our primary metric for performance evaluation is the Area Under the Receiver Operating Characteristic curve (AUROC).
Formally, given a set of true positive rates (TPR) and false positive rates (FPR) across different thresholds, the AUROC can be represented as:
 \begin{equation}
 \text{AUROC} = \int_{0}^{1} \text{TPR}(t) \, dt,
 \end{equation}
 where $t$ denotes varying threshold values. 
It provides a comprehensive view of performance across all possible thresholds, making it threshold-independent. 
This makes the metric both interpretable and insightful, offering a clearer picture of the model's discriminating capabilities.


\subsection{Baselines}\label{sec_baselines_and_models}

Our evaluation is benchmarked against a diverse range of zero-shot machine-generated text detection techniques.
And a supervised baseline is also included to demonstrate the advantages of our zero-shot method:

\begin{itemize}[leftmargin=*]
\item \textbf{Log $p(x)$}~\cite{solaiman2019release}: Utilizes the source model's average token-wise log probability to gauge code naturalness. Machine-generated code tends to have a higher score.
\item \textbf{Entropy}~\cite{gehrmann2019gltr}: Interprets high average entropy in the model’s predictive distribution as indicative of machine generation.
\item \textbf{(Log-)~Rank}~\cite{gehrmann2019gltr,ippolito2020automatic}: The average observed rank or log rank of each token in the LLM prediction, with machine-generated passages typically showing smaller average values.
\item \textbf{DetectGPT}~\cite{mitchell2023detectgpt}: Leverages the log probability of the original code and its perturbed variants to compute the perturbation discrepancy gap. 
\item \textbf{DetectLLM}~\cite{su2023detectllm}: Introduces two methods, one blends log likelihood with log rank to compute \textbf{LRR}, and the other improves DetectGPT by incorporating the \textbf{NPR} score.
\item \textbf{GPTSniffer}~\cite{nguyen2023this}: A supervised baseline that trains CodeBERT~\cite{feng2020codebert} to predict the authorship of a given code snippet. Following OpenAI's RoBERTa-based~\cite{liu2019roberta} GPT detector\footnote{\url{https://github.com/openai/gpt-2-output-dataset/tree/master/detector}}, we train the model on a combination of 1000 samples generated by each model at each setting.
\end{itemize}

\subsection{Experimental Setup}

For code generation with different models, we adopted the top-$p$ sampling strategy~\cite{holtzman2019curious} with $p$=$0.95$ following~\cite{chen2021evaluating}.  We explored two temperature settings, $T$=$0.2$ and $T$=$1.0$, as discussed in Section~\ref{sec_experimental_setup_empirical_analysis}. The maximum length constraint for generated code was set at $128$ tokens. With respect to the perturbation-specific hyperparameters, a grid search on a held-out set from the \textit{CodeSearchNet} dataset, using the SantaCoder model~\cite{allal2023santacoder}, revealed optimal values. Consequently, we set $\alpha$ and $\beta$ to $0.5$, while $\lambda_{\text{spaces}}$=3 and $\lambda_{\text{newlines}}$=2. For all experiments, we maintained a consistent configuration of generating $50$ perturbations. 

For the DetectGPT and DetectLLM, which involve an LLM in restoring perturbed code, we utilized the CodeT5+~(770M) model~\cite{wang2023codet5}. And as for the supervised baseline GPTSniffer, we trained the CodeBERT model for 5 epochs with a batch size of 16 and a learning rate of 2e-5 with AdamW optimizer.
All experiments are conducted on 2 NVIDIA RTX 4090 GPUs with 24GB memory.


\begin{table*}[t]
  \centering
    \caption{Performance (AUROC) of various detection methods. Statistical significance $p<0.001$.}
  \resizebox{\textwidth}{!}{
  \begin{tabular}{c|c|cccc@{}c@{}cc@{}c@{}c}
  \toprule
  \multirow{2}{*}{\textbf{Dataset}} & \multirow{2}{*}{\textbf{Code LLM}} & \multicolumn{8}{c}{\textbf{Detection Methods}} \\
  \cline{3-11}
  & & $\log p(x)$ & Entropy & Rank & Log Rank & \;DetectGPT & LRR & NPR & \;GPTSniffer & \;\approach \\
  \midrule
  \multirow{6}{*}{\centering \specialcell{CodeSearchNet\\ ($T=0.2$)}} 
  & Incoder (1.3B) & 0.9810 & 0.1102 & 0.8701 & 0.9892 & 0.4735 & 0.9693 & 0.8143 & 0.9426 & \textbf{0.9896} \\
  & Phi-1 (1.3B) & 0.7881 & 0.4114 & 0.6409 & 0.7513 & 0.7210 & 0.4020 & 0.7566 & 0.3855 & \textbf{0.8287} \\
  & StarCoder (3B) & 0.9105 & 0.2942 & 0.7585 &  0.9340 & 0.6949 & 0.9245 & 0.9015 & 0.7712 & \textbf{0.9438} \\
  & WizardCoder (3B) & 0.9079 & 0.2930 & 0.7556 & 0.9120 & 0.6450 & 0.7975 & 0.8677 & 0.7433 & \textbf{0.9345} \\
  & CodeGen2 (3.7B) & 0.7028 & 0.4411 & 0.7328 & 0.7199 & 0.6051 & 0.7997 & 0.6177 & 0.5327 & \textbf{0.8802} \\
  & CodeLlama (7B) & 0.8850 & 0.3174 & 0.7265 & 0.9016 & 0.8212 & 0.8332 & 0.5890 & 0.7496 & \textbf{0.9095} \\
  \midrule
  \multirow{6}{*}{\centering \specialcell{CodeSearchNet\\($T=1.0$)}} 
  & Incoder (1.3B) & 0.7724 & 0.4167 & 0.7797 & 0.7876 & 0.6258 & 0.7427 & 0.6801 & 0.6761 & \textbf{0.7882} \\
  & Phi-1 (1.3B) & 0.6118 & 0.4588 & 0.5709 & 0.6299 & 0.7492 & 0.4528 & 0.7912 & 0.4158 & \textbf{0.8365} \\
  & StarCoder (3B) & 0.6574 & 0.4844 & 0.6987 & 0.6822 & 0.6505 & \textbf{0.7050} & 0.6751 & 0.6299 & 0.6918 \\
  & WizardCoder (3B) & 0.8319 & 0.3363 & 0.7273 & 0.8338 & 0.5972 & 0.6965 & 0.7516 & 0.7068 & \textbf{0.8392} \\
  & CodeGen2 (3.7B) & 0.4484 & 0.6263 & 0.6584 & 0.4632 & 0.4797 & 0.5530 & 0.5208 & 0.4024 & \textbf{0.6798} \\
  & CodeLlama (7B) & 0.6463 & 0.4855 & 0.6759 & 0.6656 & 0.6423 & 0.6768 & 0.6515 & 0.6442 & \textbf{0.7239} \\
  \midrule
  \multirow{6}{*}{\centering \specialcell{The Stack \\ ($T=0.2$)}} 
  & Incoder (1.3B) & 0.9693 & 0.1516 & 0.8747 & 0.9712 & 0.6061 & 0.9638 & 0.8571 & 0.9291 & \textbf{0.9727} \\
  & Phi-1 (1.3B) & 0.8050 & 0.4318 & 0.6766 & 0.7622 & 0.7295 & 0.4022 & 0.8106 & 0.4640 & \textbf{0.8578} \\
  & StarCoder (3B) & 0.9098 & 0.3077 & 0.7843 & \textbf{0.9329} & 0.6824 & 0.9135 & 0.9233 & 0.7715 & 0.9274 \\
  & WizardCoder (3B) & 0.9026 & 0.3196 & 0.7963 & 0.9010 & 0.6385 & 0.7742 & 0.8574 & 0.7794 & \textbf{0.9243} \\
  & CodeGen2 (3.7B) & 0.7171 & 0.4051 & 0.7930 & 0.7301 & 0.5288 & 0.7604 & 0.5670 & 0.4520 & \textbf{0.8513} \\
  & CodeLlama (7B) & 0.8576 & 0.3565 & 0.7366 & 0.8793 & 0.8087 & 0.8358 & 0.5436 & 0.7619 & \textbf{0.8852} \\
  \midrule
  \multirow{6}{*}{\centering \specialcell{The Stack \\ ($T=1.0$)}} 
  & Incoder (1.3B) & 0.7310 & 0.4591 & 0.7673 & 0.7555 & 0.6124 & 0.7446 & 0.6787 & 0.6846 & \textbf{0.7833} \\
  & Phi-1 (1.3B) & 0.7841 & 0.4205 & 0.6666 & 0.7475 & 0.6718 & 0.4106 & 0.7755 & 0.4984 & \textbf{0.8376} \\
  & StarCoder (3B) & 0.6333 & 0.5025 & 0.7010 & 0.6609 & 0.5896 & 0.7080 & 0.6638 & \textbf{0.7243} & 0.6890 \\
  & WizardCoder (3B) & 0.8293 & 0.3459 & 0.7484 & 0.8223 & 0.6377 & 0.6436 & 0.7929 & 0.7766 & \textbf{0.8384} \\
  & CodeGen2 (3.7B) & 0.4816 & 0.6046 & 0.5631 & 0.4956 & 0.4337 & 0.5740 & 0.5178 & 0.4265 & \textbf{0.6595} \\
  & CodeLlama (7B) & 0.5929 & 0.5260 & 0.6451 & 0.6091 & 0.6116 & 0.6365 & 0.6226 & \textbf{0.7494} & 0.6660 \\  
  \midrule
   Average & - & 0.7649 & 0.3961 & 0.7228 & 0.7724 & 0.6357 & 0.7050 & 0.7178 & 0.6507 & \textbf{0.8308} \\
  \bottomrule
  \end{tabular}
  }
  \label{tab:comparison}
\end{table*}

\subsection{Detection Performance (RQ1)}

Table~\ref{tab:comparison} delineates the results of various methods. According to the results, \approach consistently outperforms baseline methods. Compared to the strongest baseline Log Rank, our method achieves an average relative improvement of 7.6\% in AUROC. In an impressive 20 of 24 combinations of dataset and model, our method provides the most accurate performance, which underscores its robustness across a variety of generative models, ranging from the 1.3 billion parameter InCoder to the 7 billion parameter CodeLlama. 

We also repeated the experiments 10 times and employed a Wilcoxon rank sum test to assess the statistical significance of the performance differences between the methods. Results show that the performance superiority of our method was statistically significant, with p-values less than 0.001.
The high AUROC scores achieved across these diverse settings confirm the method's superior capability to generalize and reliably differentiate between machine-generated and human-written code.

We can observe that the challenge of detection notably increases at a temperature setting of $T$=$1.0$ than $T$=$0.2$. This is possibly due to the higher randomness at this temperature, where models are likely to generate outputs with greater diversity in styles. Despite these increased difficulties, the proposed method maintains its leading position in detection accuracy. 

It is worth mentioning that our zero-shot framework often outperforms the supervised GPTSniffer, highlighting the challenges of detecting machine-generated code with training data-dependent supervised models. The stable performance across various generation settings showcases our method's advanced detection capabilities. This makes it an effective solution in practical applications.

\begin{figure*}[htp]
  \centering
  
  \begin{subfigure}[b]{0.48\textwidth}
      \centering
      \begin{codebox}
      \includegraphics[page=1, width=0.95\textwidth]{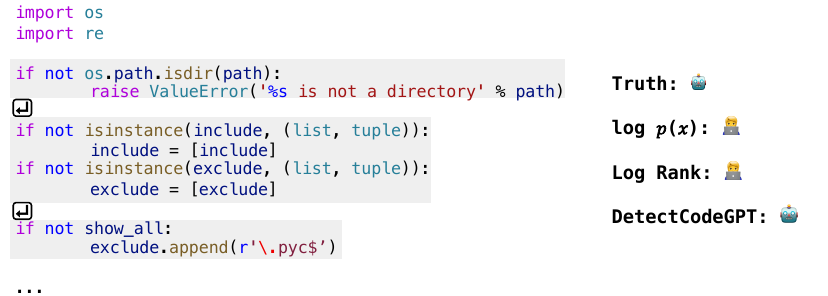}
      \end{codebox}
      \caption{Example 1}
      \label{fig:case1}
  \end{subfigure}
  \hspace{0.3cm}
  \begin{subfigure}[b]{0.48\textwidth}
      \centering
      \begin{codebox}
      \includegraphics[page=3, width=0.95\textwidth]{figures/cases.pdf}
      \end{codebox}
      \caption{Example 2}
      \label{fig:case2}
  \end{subfigure}
  
  \vspace{1em}
  
  \begin{subfigure}[b]{0.48\textwidth}
      \centering
      \begin{codebox}
      \includegraphics[page=6, width=0.95\textwidth]{figures/cases.pdf}
      \end{codebox}
      \caption{Example 3}
      \label{fig:case3}
  \end{subfigure}
  \hspace{0.3cm}
  \begin{subfigure}[b]{0.48\textwidth}
      \centering
      \begin{codebox}
      \includegraphics[page=4, width=0.95\textwidth]{figures/cases.pdf}
      \end{codebox}
      \caption{Example 4}
      \label{fig:case4}
  \end{subfigure}
  
  \caption{Examples of machine- and human-authored code snippets with corresponding predictions.}
  \vspace{-0.5cm}
  \label{fig:case_study}
\end{figure*}

To further illustrate the effectiveness of our approach, we present some representative examples by \approach and two most competitive baselines in Figure~\ref{fig:case_study}, using decision thresholds $\epsilon$ set at the mean scores of all code snippets.

Examples 1 is a machine-generated code snippet. \approach can correctly detect it, while the baselines fail. 
In this example, the first and last ``if'' statements are separated from the rest of the code with newlines, and the two ``if'' statements in the middle are modularized together since they have similar functionalities. 
Such modularization and separation of code blocks are captured by \approach thanks to the stylized perturbation.
Example 2 and 3 are human-written code that are misclassified by other baselines, but \approach correctly identifies them. In Example 2, we can observe that the code blocks are sometimes separated with newlines (e.g., lines 5-7), but sometimes not (as seen with the rest of the code). In Example 3, although the code blocks are well separated with newlines, the human author omitted the spaces between operators ``//'' and ``/'', but add spaces between ``*'' and ``+''. Such freely and randomly stylized code reveals the inherent randomness in human coding habits. 
Baselines, relying solely on token-wise conditional distributions, struggle to capture the coding style's randomness, whereas \approach effectively utilizes style information to discern between machine-generated and human-written code, demonstrating its prowess in detection through stylized code perturbation.


However, there are also cases where \approach fails. Example 4 is a human-written code misclassified as machine-generated by all the approaches. We can observe that this code snippet is well-structured with newlines and spaces. Its resemblance to machine-generated code is striking, posing a significant challenge for distinction. This example highlights the difficulty of detecting machine-generated code among well-structured human-written code with standard coding styles.

\subsection{Ablation Study (RQ2)}

\begin{table}[t]
  \centering
  \caption{Performance of different perturbation strategies}
  \label{tab:ablation_study}
  \resizebox{0.9\linewidth}{!}{
    \begin{tabular}{c@{}c@{}c@{}c@{}c}
      \toprule
      Perturb. Type\;\; & MLM\; &\; Newline\; & Space &\; Newline\&Space \\
      \midrule
      $T=0.2$ & 0.5436 & 0.8703 & 0.8639 & \textbf{0.8852} \\
      $T=1.0$ & 0.6226 & 0.6453 & 0.6504 & \textbf{0.6660} \\
      \bottomrule
    \end{tabular}
  }
  \vspace{-0.25cm}
\end{table}

In our ablation study, we compare the effectiveness of different perturbation strategies for detecting machine-generated code using the CodeLlama (7B) model on \textit{The Stack} dataset. The results summarized in Table \ref{tab:ablation_study} illuminates the comparative advantage of our stylistic perturbation approach. We observe that both newline and space perturbations independently offer substantial improvements over the traditional MLM-based (CodeT5+) perturbation technique as in DetectGPT and DetectLLM~\cite{mitchell2023detectgpt,su2023detectllm} for natural language. Also, the combination of newline and space perturbations further enhances the detection performance, with the highest AUROC score of 0.8852 at $T=0.2$ and 0.6660 at $T=1.0$. The consistent outperformance of our combined perturbation strategy across both temperature settings affirms its potential as a robust solution for detecting machine-authored code.

\subsection{Impact of Perturbation Count (RQ3)}\label{sec_perturbation_count}

To gauge the impact of perturbation count on the efficacy of our method, we conduct experiments with varying numbers of perturbations.

\begin{table}[ht]
  \centering
  \caption{Impact of varying the number of perturbations} 
  \resizebox{\linewidth}{!}{
  \begin{tabular}{cccccc}
    \toprule
    \#Perturbations & 10 & 20 & 50 & 100 & 200 \\
    \midrule
    \(T=0.2\) & 0.6537 & 0.8825 & 0.8852 & \textbf{0.8855} & 0.8846 \\
    \(T=1.0\) & 0.5558 & 0.6584 & 0.6660 & 0.6660 & \textbf{0.6662} \\
    \bottomrule
  \end{tabular}
  }
  \label{tab:num_perturbations}
  \vspace{-0.25cm}
\end{table}

Results in Table \ref{tab:num_perturbations} reveal a rapid ascent in the AUROC score as the number of perturbations increases from 10 to 20, underscoring the efficiency of our perturbation approach. Notably, an increase to 20 perturbations already yields robust detection performance, with further increments leading to diminishing improvements. This suggests that our method requires a relatively small number of perturbations to effectively discern between human- and machine-authored code. This implies that our method is not only effective but also efficient. 

\subsection{Performance of Cross-Model Code Detection (RQ4)}\label{sec_cross_model}

\begin{figure}[h]
  \centering
  \vspace{-0.2cm}
  \includegraphics[width=0.7\linewidth, trim=0 0 0 20 clip]{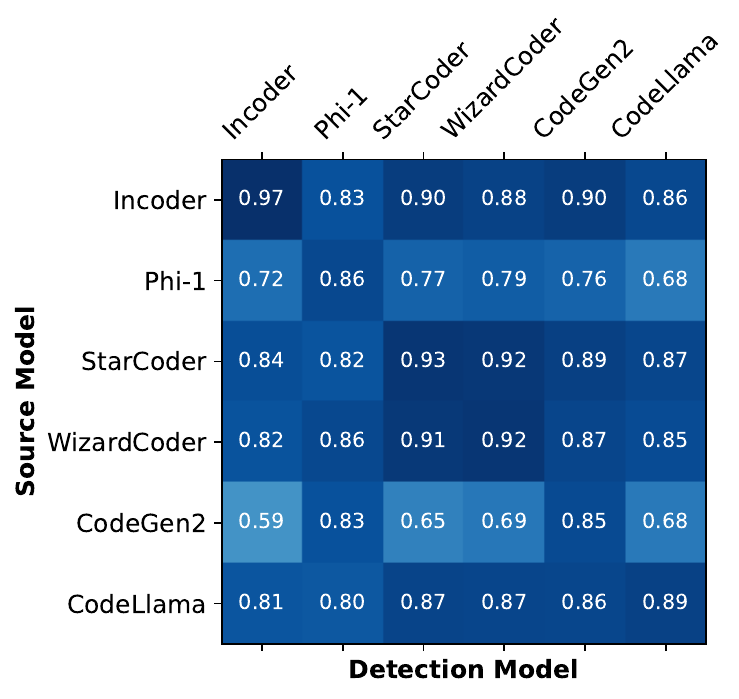}
  \vspace{-0.15cm}
  \caption{Cross-model detection performance} 
  \vspace{-0.15cm}
  \label{fig:cross_model}
\end{figure}

In previous sections, we primarily assess the efficacy of \approach within a white-box framework, where we can get access to the logits of the original code generation model, as defined in Section~\ref{sec_problem_definition}.
In real-world scenarios, accessing the original model for code generation detection is often impractical, so we conduct cross-model detection experiments, where we use other LLMs as surrogate models to compute the naturalness scores.

The evaluation results on \emph{The Stack} dataset at $T=0.2$ presented in Figure~\ref{fig:cross_model} highlight \approach's adaptability in cross-model detection. While the algorithm excels in the white-box setting, its performance endures with only a slight reduction 
in cross-model application. For instance, StarCoder, when detecting code generated by WizardCoder and CodeLlama, yields AUROC scores of 0.92 and 0.87, respectively, compared to an AUROC of 0.93 when detecting its own output. We can also notice a performance decrease when detecting CodeGen2's output. This is possibly due to the fact that CodeGen2 is trained on a more diverse dataset containing more natural language text~\cite{nijkamp2023codegen2}. However, Phi-1 demonstrates a relative proficiency with a score of 0.83 to detect CodeGen2's output, which implies an ensemble of diverse detection models may enhance the system's robustness, as suggested in~\cite{mireshghallah2023smaller}.


These results indicate that \approach is a model-free method that is robust against model discrepancies, making it a viable solution for real-world applications where the source model could be unknown or inaccessible.

\section{Discussion}\label{sec_discussion}


\subsection{Why is \approach Effective?}\label{sec_why_effective}

We attribute the effectiveness of our \approach to the following two factors: 

\subsubsection{Preservation of Code Correctness}
The mask-and-recover perturbation in DetectGPT brings minor mistakes easily (e.g., misuse of an identifier), rendering the code non-functional and has negative impact on the naturalness score.
Such code-cracking perturbations will violate the assumption of minimal impact on the code's naturalness score if it is human-written in Section~\ref{sec_detecting_machine_generated_text}.
In contrast, inserting newlines and spaces does not affect the correctness of the code in most cases, thereby ensuring the effectiveness of our method.

\subsubsection{Emulation of Human Randomness}
As discussed in Section~\ref{sec_likelihood_rank}, human inherently exhibit less naturalness and more randomness in their use of stylistic tokens such as
spaces and newlines than machines. For example, a human programmer may freely insert whitespace, especially newlines, in the code as they deem fit, whereas a machine programmer usually tries to stylize the code in a more standardized and modularized
manner. Our proposed perturbation strategy mimics human's free usage of spaces and newlines, thereby making the perturbation more ``random'' as is desired according to Section~\ref{sec_empirical_analysis}. 

\subsection{Strength of \approach}\label{sec_strength}

Compared with existing methods, \approach eliminates the need to perturb code multiple times for each LLM and thus brings more efficiency.
Compared with supervised counterparts, \approach distinguishes itself with a zero-shot learning capability, enabling it to detect machine-generated code without the necessity for training on extensive datasets. This model-agnostic advantage means that it can be generalized across various code LLMs.

\subsection{Limitations and Future Directions}

The main limitations of our work lie in the following two aspects: Firstly, due to the computational constraints, we only focus on a set of LLMs within 7B parameters. As the landscape of LLMs rapidly evolves, incorporating a wider array of more and larger LLMs could significantly bolster the generalizability and robustness of our findings. 

Secondly, our current analysis centers exclusively on Python code, while the features of other programming languages may not be fully explored. However, based on the analysis of our method in Section~\ref{sec_strength}, we believe that our method can be effectively generalized to other languages, especially where the functionality of code won't be much affected after inserting newlines and spaces like C/C++, Java, and JavaScript.

Looking ahead to future work, we plan to further improve the effectiveness of \approach when detecting machine-generated code at higher levels of generation randomness. We note from Table~\ref{tab:comparison} that although \approach outperforms other baselines at $T$=$1.0$, there is still large room for improvement. Although ensembling multiple detection models may help improve the detection performance~\cite{mireshghallah2023smaller}, we look forward to exploring more effective perturbation strategies based on code styles to further enhance the detection efficacy.

\section{Related Work}
\subsection{Machine-Generated Text Detection}
Recently, there has been much effort in detecting machine-generated text~\cite{yang2023survey,wu2023survey}. The two main categories of detection methods are zero-shot and training-based methods, and our \approach falls into the former category, which eliminates the need for training data and brings more generalization ability.

As for zero-shot methods, they are usually based on the discrepancy between likelihood and rank information of human and machine's texts~\cite{gehrmann2019gltr,ippolito2020automatic,solaiman2019release}. Leveraging the hypothesis in DetectGPT~\cite{mitchell2023detectgpt} that machine-generated text often has a negative curvature in the log probability when the text is perturbed, many perturbation based methods have been proposed~\cite{mitchell2023detectgpt,su2023detectllm,bao2023fastdetectgpt}. These methods usually perturb the text by masking a continuous span of tokens and then recover the perturbed text using another LLM like T5~\cite{ahmad2021unified}. 
The benefit of these methods is that they are zero-shot and can be applied to any LLM without access to training data. However, the perturbation process is time-consuming and computationally expensive.
When it comes to the training-based methods, fine-tuning the RoBERTa~\cite{liu2019roberta} or T5~\cite{ahmad2021unified} model with data collected from different model families at different decoding settings is a common practice~\cite{tian2023multiscale,zhan2023g3detector,chen2023gptsentinel}. Additional information like graph structure~\cite{liu2022coco}, perplexity from proxy models~\cite{wu2023llmdet} have been shown to be helpful for detection. Moreover, techniques like adversarial training~\cite{hu2023radar} and contrastive learning~\cite{liu2022coco} have also been proposed to improve the detection performance. 
The main challenge of training-based methods is that they often lack generalization ability and require access to training data from the target model~\cite{mitchell2023detectgpt}.

\subsection{Machine-Generated Code Detection}
Research on identifying machine-generated code remains relatively scarce and is purported to be more challenging than discerning machine-generated text, according to the empirical study in \cite{pan2024assessing}. 
GPTSniffer was first proposed to detect machine-generated code with supervised CodeBERT training~\cite{nguyen2023this}. Concurrent works~\cite{yang2023zeroshot} and~\cite{xu2024detecting} also explore perturbation-based methods for detecting machine-generated code, under similar framework to DetectGPT for text~\cite{mitchell2023detectgpt}. Our approach differs from these methods in that we perform a comprehensive empirical analysis of the differences between machine- and human-authored code. Based on the insights from the analysis, we proposed a innovative stylized perturbation strategy to achieve a more efficient and effective detection method.

Another related topic revolve around code watermarking techniques, which embed unique markers into the code either during the training or generation~\cite{lee2023who,sun2023codemark,li2024resilient}. The detection of these watermarks subsequently enables the recognition of code generated by machines. It should be noted, however, that these watermarking methods are primarily designed to address issues related to code licensing and plagiarism~\cite{cox2002digital,collberg1999software}. Their reliance on modifications to the generation model renders them unsuitable for general code detection tasks.

\section{Conclusion}\label{sec_conclusion}

In this paper, we perform an in-depth analysis of the nuanced differences between machine- and human-authored code across three aspects of code including lexical diversity, conciseness, and naturalness. 
The results provide new insights that machines tend to write more concise and natural code, adhering to common programming paradigms, and the disparity is more pronounced in stylized tokens such as whitespaces that represent the syntactic segmentation of code.
Based on these insights, we have proposed a new detection method, \approach, which introduces a novel stylized perturbation strategy that is simple yet effective. The experimental results of \approach confirm its effectiveness, demonstrating its potential to help maintain the authorship and integrity of code.



\section*{Acknowledgement}
This research is supported by the National Key Research and Development Program of China (Grant No. 2023YFB4503802) and the National Natural Science Foundation of China (Grant No. 62102244).
We would like to thank the anonymous reviewers for their valuable feedback and suggestions.

\balance
\bibliographystyle{IEEEtran}
\bibliography{reference}

\newpage
\section*{Appendix}

\setcounter{section}{0}

Some results are omitted in the main text due to space constraints. We provide these additional results and analyses in this appendix.

\section{Corresponding Results for Empirical Analysis}

The top tokens from machine-authored code at $T=1.0$ are shown in Table~\ref{tab:top_tokens_tp1.0}, corresponding to the results in Table~\ref{tab:top_tokens} in the main text. Similar attention on exception handling and object-oriented programming tokens can also be observed at $T=1.0$.

\begin{table}[ht]
  \centering
  \caption{Top tokens from machine-authored code at $T=1.0$}
  \begin{tabular}{c|l}
    \toprule
    \textbf{Rank} & \textbf{Machine-Authored Tokens} \\
    \midrule
    1-10 & \textbf{. ' ( ) self : , = " if} \\
    11-20 & \textbf{return} def [ ] None in == not is import \\
    21-30 & for \{ \textbf{raise \_\_name\_\_} from \} - data else get \\
    31-40 & elif value path isinstance True + os \textbf{ValueError} key text \\
    41-50 & x type 0 1 len append replace str @ f \\
    \bottomrule
  \end{tabular}
  \label{tab:top_tokens_tp1.0}
\end{table}

\begin{figure}[ht]
  \centering
  \includegraphics[width=0.9\linewidth, trim=0 20 0 0 clip]{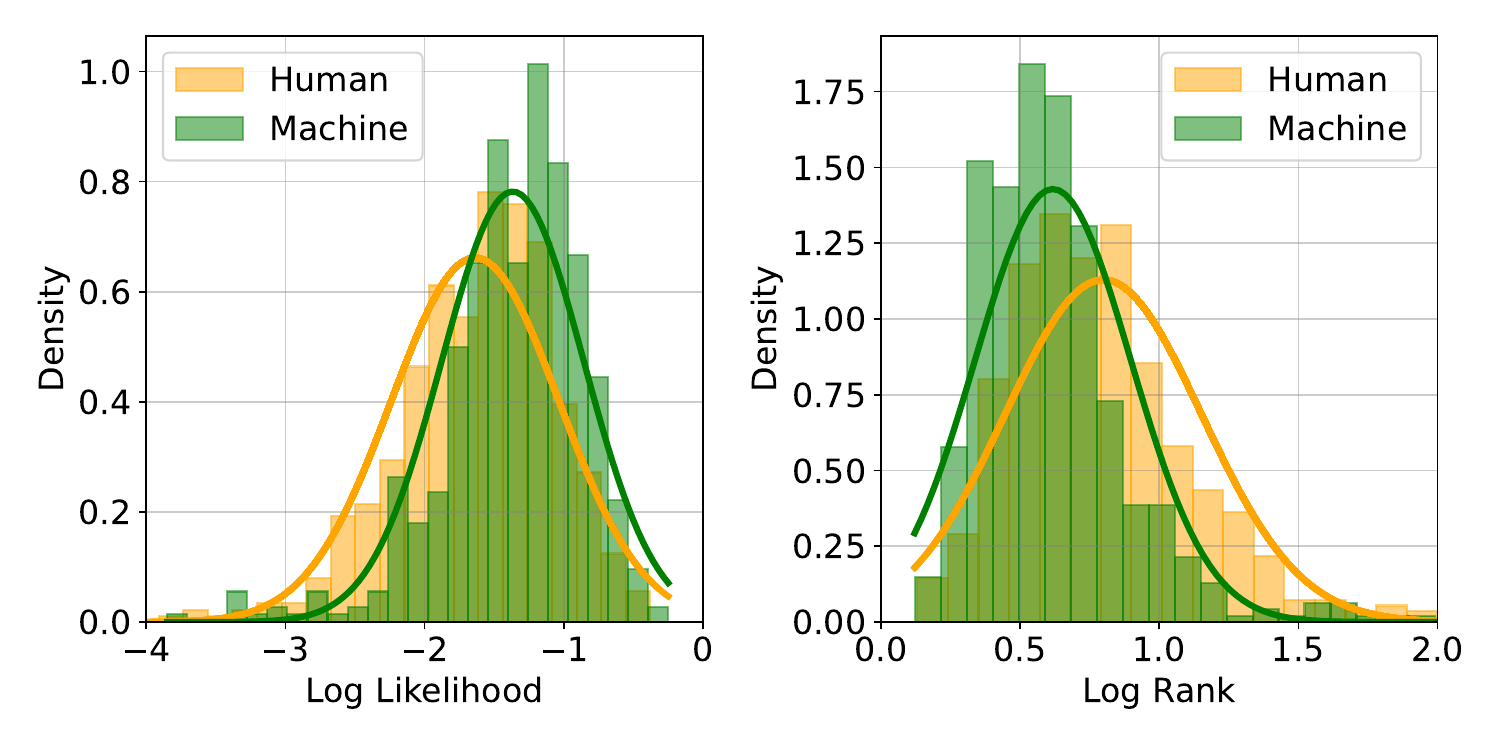}
  \vspace{-0.1cm}
  \caption{Distribution of naturalness scores with $T=1.0$}
  \label{fig:likelihood_rank_distribution_tp1.0}
  \vspace{-0.3cm}
\end{figure}

Figure~\ref{fig:likelihood_rank_distribution_tp1.0} and Table~\ref{tab:token_analysis} provide the distribution of naturalness scores and the naturalness of different categories of syntax elements, respectively. They correspond to the empirical analysis in Figure~\ref{fig:likelihood_rank_distribution} and Table~\ref{tab:token_analysis} in the main text.

We can observe from Figure~\ref{fig:likelihood_rank_distribution_tp1.0} that the trend of naturalness scores is similar to that at $T=0.2$. although the overlap between machine- and human-authored code is greater. The naturalness of different categories of syntax elements in Table~\ref{tab:token_analysis} shows that whitespace tokens is still the most effective feature as in $T=0.2$.

\begin{table}[ht]
  \centering
  \caption{The naturalness of different categories of syntax elements with $T=1.0$}
  \resizebox{\linewidth}{!}{
    \begin{tabular}{l|ccc|ccc}
      \toprule
      \multirow{2}{*}{\bf Category} & \multicolumn{3}{c|}{\bf Log Likelihood} & \multicolumn{3}{c}{\bf Log Rank} \\ \cline{2-7}
       &  Machine  &  Human & $\Delta$ &  Machine  & Human & $\Delta$ \\
       \midrule 
       keyword     & -2.025 & -2.128 & 0.103  & 0.968  & 1.053 & 0.085 \\
       identifier  & -0.787 & -0.874 & 0.087  & 0.328  & 0.378 & 0.050 \\
       literal     & -1.059 & -1.364 & 0.305  & 0.454  & 0.630 & 0.176 \\
       operator    & -1.614 & -1.835 & 0.221  & 0.733  & 0.872 & 0.139 \\
       symbol      & -0.968 & -1.639 & 0.671  & 0.321  & 0.781 & 0.460 \\
       comment     & -2.395 & -3.028 & 0.633  & 1.180  & 1.610 & 0.430 \\
       whitespace  & -2.058 & -2.740 & \textbf{0.682} & 0.946  & 1.441 & \textbf{0.495} \\
       \hline
       ALL         & -1.367 & -1.658 & 0.291  & 0.618 & 0.811 & 0.193 \\
      \bottomrule
    \end{tabular}
  }
  \label{tab:token_analysis}
  \vspace{-0.5cm}
\end{table}

\section{Cross-model Detection Performance at $T=1.0$}

\begin{figure}[ht]
  \centering
  \includegraphics[width=0.7\linewidth, trim=0 0 0 20 clip]{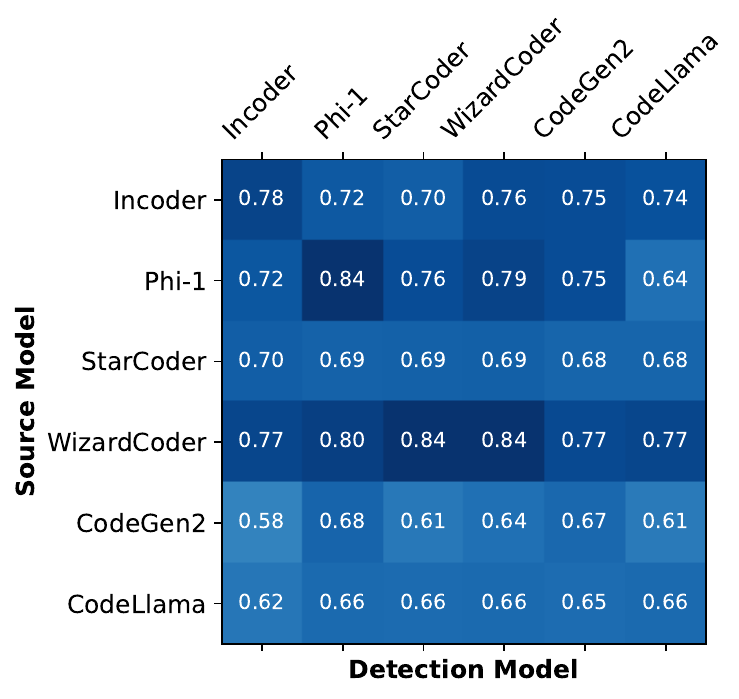}
  \caption{Cross-model detection performance at $T=1.0$}
  \label{fig:cross_model_tp1.0}
\end{figure}

Figure~\ref{fig:cross_model_tp1.0} illustrates the cross-model detection performance at $T=1.0$, corresponding to the results in Figure~\ref{fig:cross_model} in the main text. The results show that the cross-model detection performance of \approach has a similar trend across different temperatures.


\section{Impact of Varying Length for Detection}

\begin{figure}[ht]
  \centering
  \begin{subfigure}{0.4\textwidth}
  \centering
  \includegraphics[width=\textwidth]{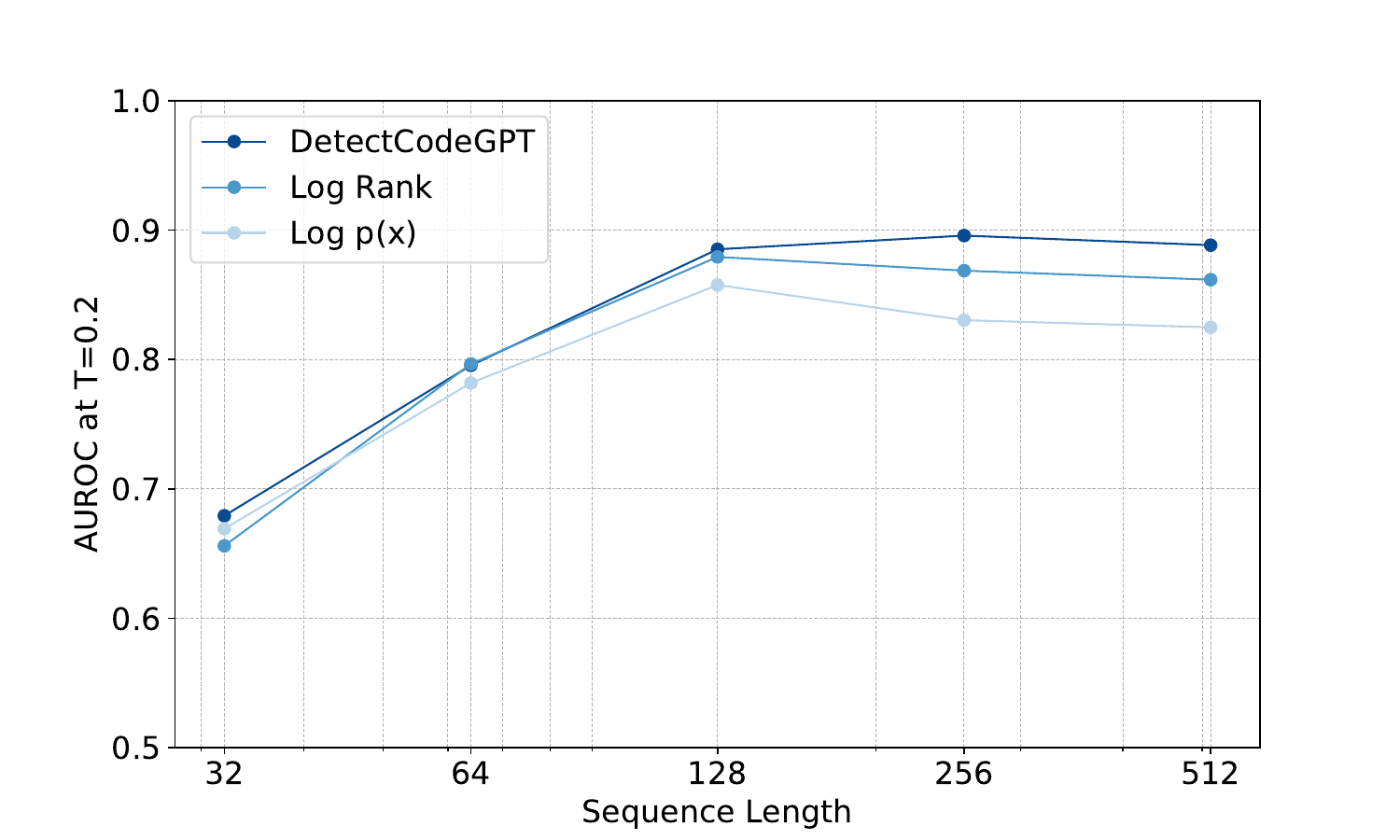}
  \caption{AUROC at $T=0.2$}
  \label{fig:varying_length_tp0.2}
  \end{subfigure}
  \hfill
  \begin{subfigure}{0.4\textwidth}
  \centering
  \includegraphics[width=\textwidth]{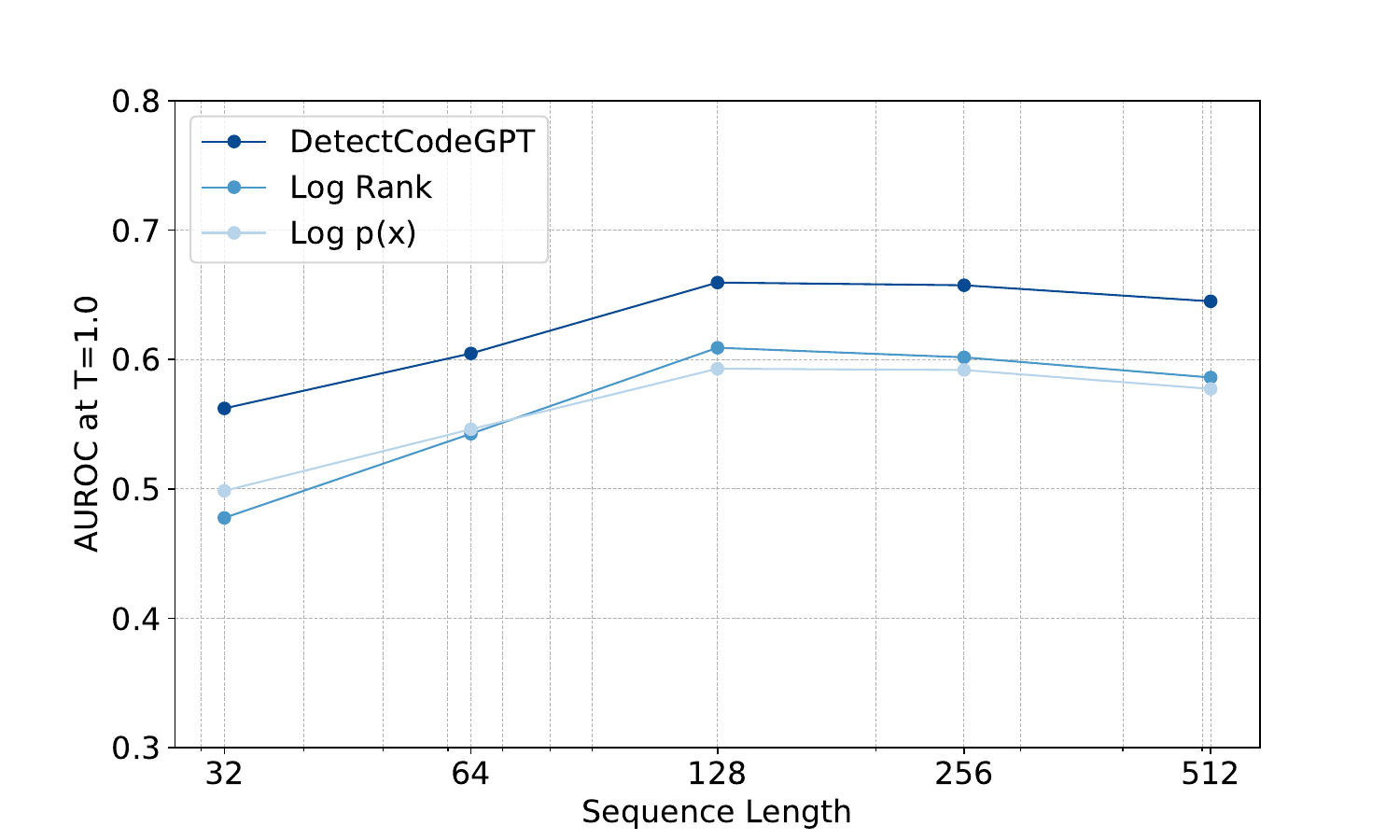}
  \caption{AUROC at $T=1.0$}
  \label{fig:token_dist_tp1.0}
  \end{subfigure}
  \caption{AUROC with different code trimming length}
  \label{fig:varying_length}
  \vspace{-0.5cm}
\end{figure}

The maximum length of code is trimmed to 128 tokens in our experiments. Figure~\ref{fig:varying_length} illustrates the trend in detection performance as the length of code snippets varies across different temperatures, compared against two of the most competitive baselines. This experiment uses the Stack dataset with CodeLlama. The results show that trimming the first 128 tokens of the snippets is sufficient to maintain high detection performance.

\end{document}